\documentclass[11pt]{article}

\usepackage[margin=1.1in]{geometry}

\usepackage{amsmath, amssymb, amsthm, mathtools}
\usepackage{bm}
\usepackage{mathrsfs}
\usepackage{thmtools}
\usepackage{thm-restate}

\NewDocumentCommand{\lbsPriorUpper}{s}{\storedLbsPriorUpper*}
\NewDocumentCommand{\lbsPriorLower}{s}{\storedLbsPriorLower*}
\NewDocumentCommand{\lbsImprovedSparsifier}{s}{\storedLbsImprovedSparsifier*}
\NewDocumentCommand{\lbsCommunication}{s}{\storedLbsCommunication*}
\NewDocumentCommand{\lbsStreamingSparsifier}{s}{\storedLbsStreamingSparsifier*}
\NewDocumentCommand{\lbsStreamingLoadBalancing}{s}{\storedLbsStreamingLoadBalancing*}
\NewDocumentCommand{\lbsLinearSizeLower}{s}{\storedLbsLinearSizeLower*}

\usepackage{appendix}
\usepackage[T1]{fontenc}
\usepackage{lmodern}
\usepackage{microtype}

\usepackage{enumitem}
\usepackage{xcolor}
\usepackage[colorlinks=true, linkcolor=blue!60!black, citecolor=blue!60!black, urlcolor=blue!60!black]{hyperref}
\usepackage[nameinlink, capitalize]{cleveref}

\usepackage{tikz}
\usetikzlibrary{arrows.meta, positioning, calc}
\usepackage[most]{tcolorbox}
\usepackage{xcolor}
\usepackage{amsmath,amssymb}

\definecolor{pastelblue}{RGB}{230,242,255}

\allowdisplaybreaks

\usepackage{amsthm}

\usepackage[nameinlink,capitalize]{cleveref}

\theoremstyle{plain}

\newtheorem{theorem}{Theorem}
\newtheorem{lemma}[theorem]{Lemma}
\newtheorem*{lemma*}{Lemma}     

\newtheorem*{theorem*}{Theorem}     

\newtheorem{claim}[theorem]{Claim}

\newtheorem{conjecture}[theorem]{Conjecture}

\newtheorem{proposition}[theorem]{Proposition}

\theoremstyle{definition}
\newtheorem{definition}[theorem]{Definition}

\theoremstyle{remark}
\newtheorem{remark}[theorem]{Remark}

\usepackage[margin=1.1in]{geometry}
\usepackage{amsmath,amssymb,amsthm,mathtools}
\usepackage{enumitem}
\usepackage{hyperref}
\usepackage{microtype}
\usepackage{xcolor}

\hypersetup{
  colorlinks=true,
  linkcolor=blue!50!black,
  citecolor=blue!50!black,
  urlcolor=blue!50!black
}

\usepackage{todonotes}

\definecolor{garytodo}{HTML}{EAF3FF}
\definecolor{sepidehtodo}{HTML}{F2ECFA}
\definecolor{jakubtodo}{HTML}{FFF3E4}

\definecolor{conjecturebackground}{HTML}{FFFBEA}
\definecolor{conjectureborder}{HTML}{D8C878}

\newsavebox{\conjecturestatementbox}

\newenvironment{yellowboxedconjecture}
{%
    \par\medskip
    \begingroup
    \setlength{\fboxsep}{10pt}%
    \setlength{\fboxrule}{0.6pt}%
    \begin{lrbox}{\conjecturestatementbox}%
    \begin{minipage}{\dimexpr\linewidth-2\fboxsep-2\fboxrule\relax}%
    \color{black}%
}
{%
    \end{minipage}%
    \end{lrbox}%
    \noindent
    \fcolorbox{conjectureborder}{conjecturebackground}
        {\usebox{\conjecturestatementbox}}%
    \endgroup
    \par\medskip
}

\newcommand{\polylog}{\operatorname{polylog}}
\newcommand{\floor}[1]{\left\lfloor #1 \right\rfloor}
\newcommand{\ceil}[1]{\left\lceil #1 \right\rceil}

\newcommand{\Oish}{\widetilde{O}}
\newcommand{\Omegaish}{\widetilde{\Omega}}

\definecolor{redtheorembackground}{HTML}{FCF3F3}
\definecolor{redtheoremborder}{HTML}{D9AAAA}

\newsavebox{\redtheoremstatementbox}

\newenvironment{redboxedtheorem}
{%
    \par\medskip
    \begingroup
    \setlength{\fboxsep}{10pt}%
    \setlength{\fboxrule}{0.6pt}%
    \begin{lrbox}{\redtheoremstatementbox}%
    \begin{minipage}{\dimexpr\linewidth-2\fboxsep-2\fboxrule\relax}%
    \color{black}%
}
{%
    \end{minipage}%
    \end{lrbox}%
    \noindent
    \fcolorbox{redtheoremborder}{redtheorembackground}
        {\usebox{\redtheoremstatementbox}}%
    \endgroup
    \par\medskip
}

\definecolor{theorembackground}{HTML}{F2F8F2}
\definecolor{theoremborder}{HTML}{A9CDAE}

\newsavebox{\theoremstatementbox}

\newenvironment{boxedtheorem}
{%
    \par\medskip
    \begingroup
    \setlength{\fboxsep}{10pt}%
    \setlength{\fboxrule}{0.6pt}%
    \begin{lrbox}{\theoremstatementbox}%
    \begin{minipage}{\dimexpr\linewidth-2\fboxsep-2\fboxrule\relax}%
    \color{black}%
}
{%
    \end{minipage}%
    \end{lrbox}%
    \noindent
    \fcolorbox{theoremborder}{theorembackground}
        {\usebox{\theoremstatementbox}}%
    \endgroup
    \par\medskip
}

\definecolor{definitionbackground}{HTML}{F1F8FC}
\definecolor{definitionborder}{HTML}{B7D9EC}
\newsavebox{\definitionbox}
\newenvironment{boxeddefinition}
{%
    \par\medskip
    \begingroup
    \setlength{\fboxsep}{10pt}%
    \setlength{\fboxrule}{0.6pt}%
    \begin{lrbox}{\definitionbox}%
    \begin{minipage}{\dimexpr\linewidth-2\fboxsep-2\fboxrule\relax}%
}
{%
    \end{minipage}%
    \end{lrbox}%
    \noindent
    \fcolorbox{definitionborder}{definitionbackground}{\usebox{\definitionbox}}%
    \endgroup
    \par\medskip
}

\title{
Improved Sparsification for Load Balancing
}
\author{
  \begin{tabular}{c@{\hspace{3em}}c@{\hspace{3em}}c}
    Gary Hoppenworth\thanks{Work done while an intern at Microsoft Research.}
      & Sepideh Mahabadi & Jakub Tarnawski \\
      \normalsize TTIC & 
    \normalsize Microsoft Research & \normalsize Microsoft Research \\
    \normalsize\texttt{garytho@umich.edu} & \normalsize\texttt{smahabadi@microsoft.com}  &
    \normalsize \texttt{jatarnaw@microsoft.com}
  \end{tabular}
}
\date{}

\begin{document}

\maketitle

\begin{abstract}

In the load-balancing problem, we have a bipartite graph $G = (C \cup S, E)$ with clients $C$ on the left and servers $S$ on the right, and the goal is to assign each client to an adjacent server, while minimizing the maximum load on any one server. 
Load balancing is well-studied in the online model, where clients arrive one at a time and we have to assign them to a server immediately. We study load balancing in two other models where we have incomplete information about the input: the two-party communication model and the streaming model. We show:
\begin{itemize}
    \item There exists a single-pass streaming algorithm that uses $\Oish(n)$ space and computes an $\Oish(n^{1/3})$-approximation to load balancing. This improves the prior $O(n^{1/2})$-approximation due to Konrad and Ros\'en [ICALP'13].
    \item There exists a one-way communication protocol that uses $\Oish(n)$ bits of communication and obtains an $n^{1/3-\varepsilon_0}$-approximation to load balancing, for some $\varepsilon_0 > 0$. This improves the previous $O(n^{1/3})$-approximation due to Konrad and Ros\'en. 
\end{itemize}
For both of these problems, there is an $n^{1/4-o(1)}$ lower bound on the best possible approximation factor, due to Assadi, Bernstein, Langley, Lau, and Wang [SODA'25]. Thus we make meaningful progress on closing the approximation-factor gap for both problems. 

At a technical level, we obtain our results by giving improved constructions for \emph{load-balancing sparsifiers}, sparse subgraphs $H \subseteq G$ that preserve the most important client-server feasibility pairs in $G$.

\end{abstract}

\newcommand{\load}{\operatorname{load}}
\newcommand{\OPTload}{\operatorname{\textsc{opt}}}

\section{Introduction}

A common problem in networking and distributed systems is to assign users (or jobs) to servers without placing too much load on any one server. This is known as the load-balancing problem.

\begin{boxeddefinition}
\begin{definition}[Load balancing]
We are given a set of clients \(C\)  and a disjoint set of servers \(S\), and, for each
client \(c\in C\), a nonempty set \(S_c\subseteq S\) of feasible servers. A
feasible assignment is a map \(\sigma\colon C\to S\) satisfying
\(\sigma(c)\in S_c\) for every \(c\in C\). For an assignment \(\sigma\), the
\emph{load} of a server \(s\in S\) is
\(
    \load_\sigma(s)
    \mathrel{:=}
    \bigl|\{c\in C:\sigma(c)=s\}\bigr|.
\)
The \emph{load-balancing problem} is to find a feasible assignment minimizing
\[
    \max_{s\in S}\load_\sigma(s).
\]
\end{definition}
\end{boxeddefinition}

Load balancing has a natural representation as a bipartite graph 
\(G=(C\cup S,E)\), with the clients \(C\) on the left, the servers \(S\) on
the right, and an edge \((c,s)\in E\) precisely when \(s\in S_c\). We call
\(G\) the \emph{assignment graph}. For any such graph \(G\), let
\(\OPTload(G)\) denote the minimum possible maximum server load among
assignments using only edges of \(G\).

An assignment graph can contain as many as \(|C||S|\) edges. When
both sides are large, this quadratic blow-up makes storing or
communicating the graph prohibitively expensive. It is therefore natural to
ask whether we can store a \emph{sparse} subgraph \(H\subseteq G\) which preserves only the ``most important'' feasibility edges in $G$. 
This motivates the notion of a \emph{load-balancing sparsifier.}


\begin{boxeddefinition}

\begin{definition}
[Load-balancing sparsifier~\cite{konrad2016approximating, assadi2025streaming}]
    Let \(G=(C\cup S,E)\) be an assignment graph, and let
\(\alpha\geq 1\). A subgraph \(H\subseteq G\) is an
\(\alpha\)-load-balancing sparsifier, or \(\alpha\)-LBS, if for every
\(X\subseteq C\),
\[
    \OPTload\bigl(H[X\cup S]\bigr)
    \leq
    \alpha\cdot \OPTload\bigl(G[X\cup S]\bigr).
\]
\end{definition}

\end{boxeddefinition}



The strength of load-balancing sparsifiers lies in  their ability to approximate $\OPTload(G[X \cup S])$ for every subset of clients $X \subseteq C$. 
This robustness  makes load-balancing sparsifiers particularly useful for solving  load balancing in settings where we have incomplete information about the assignment graph.

Load-balancing sparsifiers were first introduced by Konrad and Ros\'en~\cite{konrad2016approximating} as a tool for solving load balancing in the one-way communication and streaming models. In these settings, we are primarily interested in load-balancing sparsifiers of near-linear size, i.e., at most $|E(H)| = \Oish(|C|)$ edges. The work of \cite{konrad2016approximating} established the following  bounds.

\begin{yellowboxedconjecture}

\begin{restatable}[\cite{konrad2016approximating}]{theorem}{storedLbsPriorUpper}
\label{thm:prior-upper}
Every assignment graph \(G=(C\cup S,E)\) with \(n=|C|\) clients
admits
\begin{enumerate}
    \renewcommand{\labelenumi}{(\roman{enumi})}
    \item an \(O(n^{1/2})\)-load-balancing sparsifier with at most \(n\) edges; and
    \item an \(O(n^{1/3})\)-load-balancing sparsifier with at most \(2n\) edges.
\end{enumerate}
Moreover, for every integer $c > 0$, $n^{1/(c+1) - o(1)}$-load-balancing sparsifiers require  $cn$ edges.  
\end{restatable}
\end{yellowboxedconjecture}

Inspecting the bounds in \Cref{thm:prior-upper}, one might conjecture that there exists $O_c(n^{1/(c+1)})$-load-balancing sparsifiers with $cn$ edges for every integer $c \ge 1$. In a surprising turn of events, Assadi, Bernstein, Langley, Lau, and Wang~\cite{assadi2025streaming} showed that improving the approximation quality significantly beyond $n^{1/4}$ requires load-balancing sparsifiers of superlinear size.

\begin{yellowboxedconjecture}
    
\begin{restatable}[\cite{assadi2025streaming}]{theorem}{storedLbsPriorLower}
    
\label{thm:prior-lower}
\(O(n^{1/4-\varepsilon})\)-load-balancing sparsifiers require 
\(
    n^{1+\Omega(\varepsilon^2)}
\)
edges for every $\varepsilon > 0$. 
\end{restatable}

\end{yellowboxedconjecture}

Theorems~\ref{thm:prior-upper} and~\ref{thm:prior-lower} leave an intriguing gap: the best approximation for $\Oish(n)$-edge load-balancing sparsifiers lies somewhere between  $n^{1/4-o(1)}$ and $O(n^{1/3})$, where $n=|C|$ is the number of clients. 
Our first main result makes progress on the upper-bound side of this gap.

\begin{boxedtheorem}

\begin{restatable}{theorem}{storedLbsImprovedSparsifier}
\label{thm:improved-sparsifier}
There is an absolute constant \(\varepsilon_0>0\) such that every
assignment graph \(G=(C\cup S,E)\) with \(n=|C|\) clients admits an
\(O(n^{1/3-\varepsilon_0})\)-load-balancing sparsifier with \(O(n\log^3 n)\) edges.
\end{restatable}
\end{boxedtheorem}

\noindent 
The value obtained by our analysis is small, roughly 
\(\varepsilon_0\approx 0.007\). Nevertheless, our improvement is
polynomial, and it is the first improvement over the \(O(n^{1/3})\) bound
since 2013. We establish this bound by developing a new combinatorial analysis of \emph{matching contractors}, the dual objects of load-balancing sparsifiers that were introduced in ~\cite{assadi2025streaming}. (See Section \ref{subsec:LBS-over} for a proof overview.)


As an immediate corollary of \Cref{thm:improved-sparsifier} and the reductions of~\cite{konrad2016approximating, assadi2025streaming}, we obtain an improved upper bound on the one-way communication complexity of load balancing. 
In this model, the assignment graph is divided between two players. The first player sends a message to the second player, and the second player  uses the message  to compute a load-balancing solution.\footnote{See \Cref{def:comm} for a formal definition of the two-party one-way communication model.}


\begin{boxedtheorem}

\begin{restatable}{corollary}{storedLbsCommunication}
\label{cor:communication}
In the one-way communication model, there is a deterministic protocol using
\(\widetilde{O}(n)\) bits of communication that obtains an
\(O(n^{1/3-\varepsilon_0})\)-approximation for load balancing.
\end{restatable}
\end{boxedtheorem}

\noindent 
Prior reductions in \cite{assadi2025streaming}  show that the
optimal approximation achievable by an $\Oish(n)$-bit one-way communication protocol is asymptotically
\emph{equal} to the optimal approximation achievable 
by an $\Oish(n)$-size load-balancing sparsifier. 
Thus, in the one-way communication model, the optimal approximation for load balancing now lies between
\(n^{1/4-o(1)}\) and \(n^{1/3-\varepsilon_0}\).

We also study streaming algorithms for computing load-balancing sparsifiers.
In the streaming model, edges of the assignment graph arrive in a stream and we must solve load balancing  using limited local memory.\footnote{See \Cref{def:streaming} for a formal definition of the graph streaming model. } 
We show how to adapt the existential upper bounds in \Cref{thm:prior-upper} due to~\cite{konrad2016approximating} into a streaming algorithm, and obtain our second main result.

\begin{boxedtheorem}
    
\begin{restatable}{theorem}{storedLbsStreamingSparsifier}
\label{thm:streaming-sparsifier}
\label{thm:streaming-sparsifier-restatement}
There is a single-pass streaming algorithm that uses $\Oish(n)$ space and computes an $\Oish(n^{1/3})$-load-balancing sparsifier with $\Oish(n)$ edges. 
\end{restatable}
\end{boxedtheorem}

As an immediate consequence, we obtain
an improved streaming algorithm for load balancing.

\begin{boxedtheorem}

\begin{restatable}{corollary}{storedLbsStreamingLoadBalancing}
\label{cor:streaming-load-balancing}
There is a single-pass streaming algorithm that uses 
\(\widetilde{O}(n)\) space and computes an 
\(\widetilde{O}(n^{1/3})\)-approximation for load
balancing.
\end{restatable}
\end{boxedtheorem}

Previously, only an \(O(\sqrt{n})\)-approximation was known in one pass and $\Oish(n)$ space due to \cite{konrad2016approximating}. 
Conversely, the \(n^{1/4-o(1)}\) lower bound for the one-way communication
complexity of load balancing implies the same lower bound on the
approximability of load balancing in the semi-streaming model~\cite{assadi2025streaming}. Thus the optimal approximation for  single-pass and $\Oish(n)$-space streaming  algorithms for load balancing lies between $n^{1/4-o(1)}$ and $\Oish(n^{1/3})$. 

\paragraph{
What are the optimal bounds for load-balancing sparsifiers?}
We now take a moment to speculate about the optimal bounds for load-balancing sparsifiers. 
Theorem~\ref{thm:prior-upper} shows that there exist 
\begin{itemize}
    \item $n$-edge $O(n^{1/2})$-load-balancing sparsifiers; and 
    \item $2n$-edge  $O(n^{1/3})$-load-balancing sparsifiers.  
\end{itemize} 
It seems reasonable to conjecture that there exist $3n$-edge $O(n^{1/4})$-load-balancing sparsifiers. Surprisingly, this is false. In fact, no constant multiple of \(n\) edges
reaches the \(n^{1/4}\) threshold.

\begin{boxedtheorem}
\begin{restatable}{theorem}{storedLbsLinearSizeLower}
\label{thm:linear-size-lower} For every  $c > 1$, 
\(O(n^{1/4+ 1/c})\)-load-balancing sparsifiers require $\Omega(cn)$ edges.
\end{restatable}
\end{boxedtheorem}
Our proof of \Cref{thm:linear-size-lower} follows the general lower-bound framework of~\cite{assadi2025streaming}, but adds one new ingredient. Taken together with Theorem~\ref{thm:improved-sparsifier}, these bounds lead us to the following conjecture, which we leave as an open problem.

\begin{redboxedtheorem}
\begin{conjecture}
Every assignment graph \(G=(C\cup S,E)\) with \(n=|C|\) clients admits an \(O(n^{1/4})\)-load-balancing sparsifier with \(O(n \log n)\) edges.
\end{conjecture}
\end{redboxedtheorem}

\paragraph{Related Work.}
The load-balancing problem can be phrased in terms of \emph{semi-matchings}, a relaxation of ordinary matchings. A semi-matching is a subgraph in which every client has degree one, while servers may have larger degree; the degree of each server is its load. Harvey, Ladner, Lov{\'a}sz, and Tamir~\cite{HarveyLadnerLovaszTamir2006} introduced this formulation and gave algorithms for finding balanced semi-matchings. Fakcharoenphol, Laekhanukit, and Nanongkai~\cite{fakcharoenphol2014faster} later developed faster exact algorithms for semi-matchings. Konrad and Ros{\'e}n studied semi-matchings in the streaming and two-party communication models~\cite{konrad2016approximating}.

Load-balancing sparsifiers are closely related to \emph{matching sparsifiers}, sparse subgraphs that approximately preserve the maximum matching size. A standard construction for matching sparsifiers is the \emph{edge-degree constrained subgraph} (EDCS). Bernstein and Stein introduced EDCSs in their work on dynamic matching~\cite{bernstein2015fully,bernstein2016faster}; subsequent work developed them as matching sparsifiers for communication, distributed, and streaming algorithms~\cite{assadi2019towards,assadi2019coresets,bernstein2020improved}.  Load-balancing sparsifiers generalize matching sparsifiers by preserving an approximate matching simultaenously for \emph{every} client subset. Indeed, if \(H\) is an \(\alpha\)-load-balancing sparsifier of \(G\), then for every \(X\subseteq L\), \(H[X\cup R]\) contains an \(\alpha\)-approximate maximum matching for \(G[X\cup R]\).

More broadly, our problem is a special case of makespan minimization on unrelated machines, studied by Lenstra, Shmoys, and Tardos~\cite{LenstraShmoysTardos1990}. More precisely, it is the unit-job case of the \emph{restricted-assignment problem}~\cite{Svensson2012}, in which each job can use only a prescribed set of machines. Azar, Naor, and Rom~\cite{AzarNaorRom1995} studied the online version, in which jobs arrive one at a time and must be assigned immediately.

\paragraph{AI Disclosure.} 

We used ChatGPT 5.5 Pro and ChatGPT 5.6 Ultra in the proof of \Cref{thm:streaming-sparsifier} in \Cref{sec:streaming-sparsifiers} and \Cref{thm:linear-size-lower} in \Cref{sec:matching-contractor-lower-bound}. We note that AI tools were not used in the proof of \Cref{thm:improved-sparsifier}. The authors verified the correctness and originality of all content including references
and take responsibility for all the content.

AI tools are rapidly changing how theoretical computer science research is conducted and disseminated. 
We believe it is useful to provide a more detailed explanation of how generative AI was used during the development of our results.
We provide this discussion in  \Cref{sec:ai_methods}.

\section{Technical Overview}

We split our technical overview into three sections:
\begin{itemize}
    \item Section \ref{subsec:LBS-over}, where we overview our load-balancing sparsifier upper bound (\cref{thm:improved-sparsifier}),
    \item Section \ref{subsec:stream-LBS}, where we overview our streaming algorithm for constructing a load-balancing sparsifier (\cref{thm:streaming-sparsifier}), and
    \item Section \ref{subsec:lower}, where we overview our load-balancing sparsifier lower bound (\Cref{thm:linear-size-lower}).
\end{itemize}

\subsection{ Load-Balancing Sparsifier Upper Bound }

\label{subsec:LBS-over}

We establish our upper bound for load-balancing sparsifiers by analyzing their dual objects, a family of graphs called matching contractors. Matching contractors are closely related to the well-studied Ruzsa--Szemer\'edi graphs and generalize them in some respects~\cite{assadi2025streaming}. 

\begin{boxeddefinition}
    \begin{definition}[Matching Contractors~\cite{assadi2025streaming}]
        An $\alpha$-matching-contractor (or $\alpha$-MC) is a bipartite graph $G = (L \cup R, E)$ that can be decomposed into a collection of matchings $M_1, \dots, M_k$ such that for each $i \in [1, k]$,
$$
|N_{G - M_i}(L(M_i))| \le |M_i| / \alpha,
$$
where $L(M_i) := V(M_i) \cap L$ and $N_{G}(X)$ denotes the neighborhood of $X$ in $G$. 
    \end{definition}
\end{boxeddefinition}

%

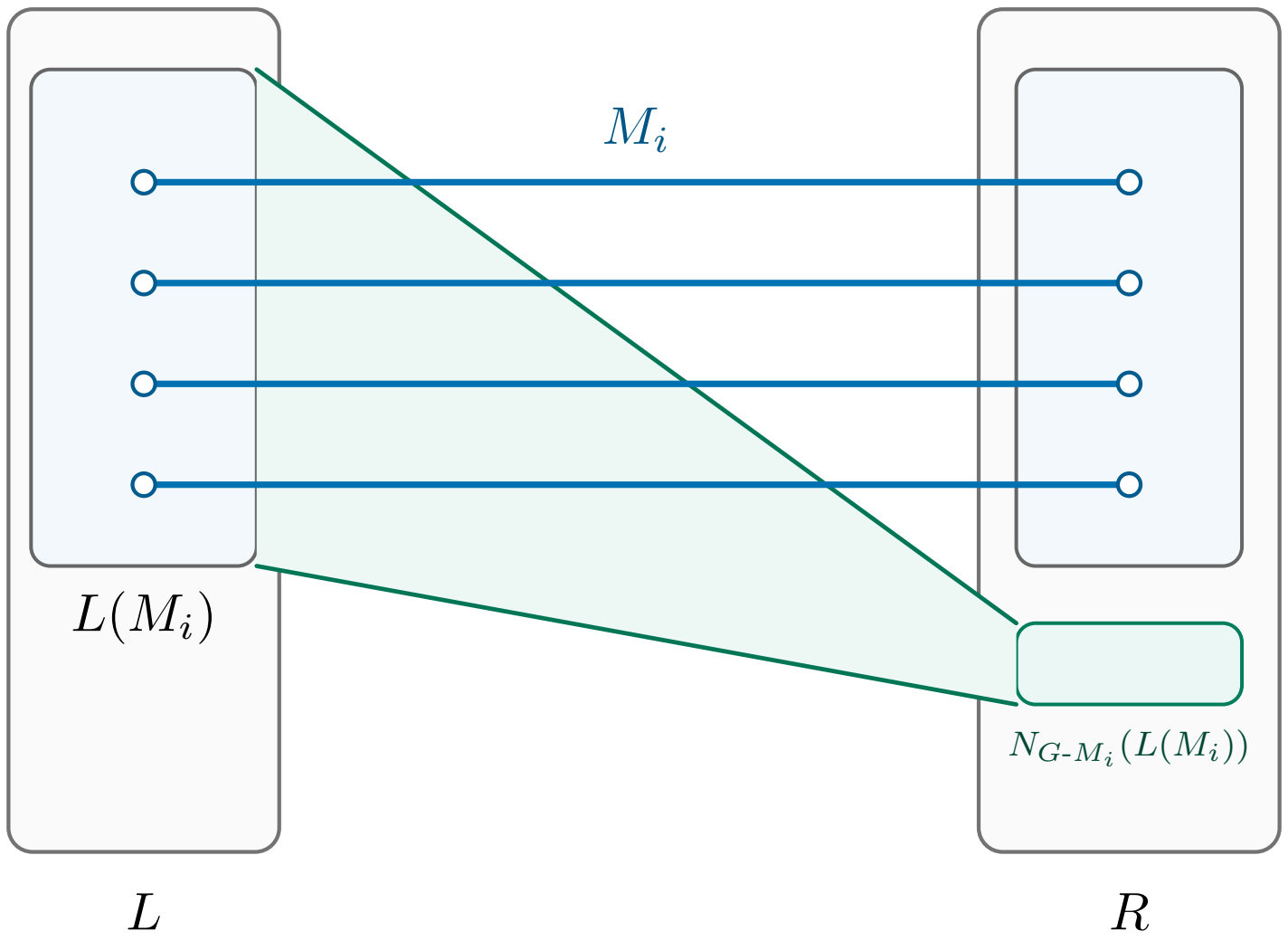
\begin{figure}[h!]
  \centering
  \definecolor{MCBlue}{HTML}{0072B2}  
  \definecolor{MCGreen}{HTML}{009E73} 

  \begin{tikzpicture}[
      x=1cm,
      y=1cm,
      scale=1.10,
      transform shape,
      line cap=round,
      line join=round,
      outer/.style={
        draw=black!55,
        fill=black!2,
        line width=.8pt,
        rounded corners=5pt
      },
      inner/.style={
        draw=black!60,
        fill=white,
        line width=.75pt,
        rounded corners=4pt
      },
      vertex/.style={
        circle,
        draw=MCBlue!80!black,
        fill=white,
        line width=.8pt,
        inner sep=0pt,
        minimum size=4.3pt
      },
      matching/.style={MCBlue, line width=1.35pt},
      residual/.style={MCGreen!75!black, line width=.9pt}
    ]

    \draw[outer] (0,0.10) rectangle (1.80,5.70);
    \draw[outer] (6.45,0.10) rectangle (8.45,5.70);
    \node[font=\small, anchor=north] at (0.90,-0.04) {$L$};
    \node[font=\small, anchor=north] at (7.45,-0.04) {$R$};

    \draw[inner, fill=MCBlue!5] (0.15,2.00) rectangle (1.65,5.30);
    \draw[inner, fill=MCBlue!5] (6.70,2.00) rectangle (8.20,5.30);
    \node[font=\small, anchor=north, inner sep=0pt]
      at (0.90,1.84) {$L(M_i)$};

    \draw[inner, draw=MCGreen!80!black, fill=MCGreen!8]
      (6.70,1.08) rectangle (8.20,1.62);
    \node[
      font=\scriptsize,
      scale=0.82,
      text=MCGreen!45!black,
      anchor=north,
      inner sep=0pt
    ] at (7.45,0.92) {$N_{G\text{-}M_i}(L(M_i))$};

    \fill[MCGreen!7]
      (1.65,5.30) -- (6.70,1.62) -- (6.70,1.08) -- (1.65,2.00) -- cycle;
    \draw[residual] (1.65,5.30) -- (6.70,1.62);
    \draw[residual] (1.65,2.00) -- (6.70,1.08);

    \foreach \j/\yy in {1/4.55,2/3.88,3/3.21,4/2.54}{
      \coordinate (left-\j)  at (0.90,\yy);
      \coordinate (right-\j) at (7.45,\yy);
      \draw[matching] (left-\j) -- (right-\j);
      \node[vertex] at (left-\j) {};
      \node[vertex] at (right-\j) {};
    }
    \node[
      font=\small,
      text=MCBlue!75!black,
      fill=white,
      inner sep=1.5pt
    ] at (4.175,4.90) {$M_i$};
  \end{tikzpicture}
 \caption{The small-neighborhood property of matching contractors. Every  neighbor of $L(M_i)$ in $G-M_i$ lies in
    a small residual neighborhood of size at most $|M_i| / \alpha$.
    }
  \label{fig:matching-contractor-small-neighborhood}
\end{figure}

Matching contractors are the canonical hard class of graphs for load-balancing sparsification. In fact, \cite{assadi2025streaming} showed that the extremal functions for load-balancing sparsifiers and matching contractors are equal, up to log factors.

\begin{yellowboxedconjecture}
    
    \begin{lemma}[ Matching Contractor Duality (Theorem 2 of \cite{assadi2025streaming})]
.\label{lem:MC_duality}
\begin{itemize}
    \item Let $\normalfont \textsf{Sparsifier}(n, \alpha)$ denote the smallest positive integer such that every assignment graph with $n$ clients admits an $\alpha$-load-balancing sparsifier on $\normalfont \textsf{Sparsifier}(n, \alpha)$ edges.

    \item Let $\normalfont \textsf{Contractor}(n, \alpha)$ denote the maximum number of edges in an $\alpha$-matching-contractor with $n$ left vertices. 
\end{itemize}
Then
$$
 1/2 \cdot  \normalfont \textsf{Contractor}(n, 2\alpha)  \le  \textsf{Sparsifier}(n, \alpha) \le O(\log n) \cdot \textsf{Contractor}(n, \alpha / 2).
$$
    \end{lemma}
\end{yellowboxedconjecture}

In light of \Cref{lem:MC_duality}, determining the optimal approximation factor of $\Oish(n)$-edge load-balancing sparsifiers is equivalent to answering the following question:
\begin{center}
    \textit{What is the smallest $\alpha(n)$ such that every $\alpha(n)$-MC with $n$ left vertices has $\Oish(n)$ edges?}
\end{center}

To make progress on this question, we develop two new technical ingredients for analyzing matching contractors: a \textbf{dispersion lemma} and a \textbf{density increment argument}.  These two ideas then combine straightforwardly to yield our improved bound.

For the remainder of this section, let $G = (L \cup R, E)$ be an $\alpha$-matching-contractor with $n$ left vertices and at least $|E| \ge n \polylog(n)$ edges. Let $M_1, \dots, M_k$ be the collection of matchings associated with $G$. To simplify our proof overview, we assume all matchings $M_i$ have exactly $s$ edges, which implies that $|E| = ks$.

\paragraph{Dispersion Lemma.} A dispersion lemma is a guarantee that certain subgraphs must be ``dispersed'' throughout the graph instead of being concentrated in any one place. For example, in the graph spanner literature, dispersion lemmas are frequently used to argue that short paths must be spread throughout the graph spanner~\cite{bodwin2025alternate}. 

We extend this perspective to matching contractors by arguing that the matchings $\{M_i\}_{i \in [k]}$ are dispersed throughout the matching contractor. At an intuitive level, the small-neighborhood condition of matching contractors 
prevents the matchings from overlapping too much. We can formalize this as follows.

\begin{boxedtheorem}
\begin{lemma}[Dispersion Lemma]
    For any two distinct matchings $M_i$ and $M_j$ of size $|M_i|=|M_j|=s$ in an $\alpha$-matching contractor $G=(L \cup R, E)$,  $$|L(M_i) \cap L(M_j)| \le s/\alpha,$$
    where $L(M_i) := V(M_i) \cap L$. 
    \label{lem:intro-dispersion}
\end{lemma}
\end{boxedtheorem}
\begin{proof}
    Let $M \subseteq M_i$ be the matching $M=\{(u, v) \in M_i \mid u \in L(M_i) \cap L(M_j)\}$. Then $V(M) \cap R \subseteq N_{G-M_j}(L(M_j))$. We conclude that
    $$
    |L(M_i) \cap L(M_j)| = |M| = |V(M) \cap R| \le |N_{G-M_j}(L(M_j))| \le |M_j| / \alpha = s/\alpha.
    $$
\end{proof}

This dispersion lemma says that no two matchings can share more than a $1/\alpha$-fraction of vertices in $L$. As it turns out, this lemma is already sufficient to recover the  state-of-the-art  $n^{1/3}$-load-balancing sparsifiers with $\Oish(n)$ edges due to~\cite{konrad2016approximating}. (See Warm-Up $\#1$ and Warm-Up $\#2$ in \Cref{sec:upper} for an accessible proof of this fact.)  

To improve beyond an $n^{1/3}$-approximation, we develop a stronger dispersion lemma. This new dispersion lemma is formalized in \Cref{sec:upper} as \Cref{lem:final_dispersion}. It is the most technically involved step in the proof of \Cref{thm:improved-sparsifier}. At a high level, we identify a new secondary structure in $G$ which guarantees that the \emph{average} pairwise overlap of the matchings $\{M_i\}_{i \in [k]}$ is significantly smaller than what is implied by \Cref{lem:intro-dispersion}.

Our stronger dispersion lemma gives a new upper bound on the approximation factor of  matching contractors that depends on the size $s$ of the matchings in $G$.

\begin{boxedtheorem}
\begin{theorem} \label{thm:intro-small-s}
    Every $\Omegaish(n^{1/6}s^{1/3})$-matching-contractor with $n$ left vertices has $\Oish(n)$ edges.
\end{theorem}
\end{boxedtheorem}

\Cref{thm:intro-small-s} is particularly strong when $s$ is small: it already gives the optimal $\Oish(n^{1/4})$-approximation factor when $s \approx \alpha$, since this implies $\alpha \approx n^{1/6}\alpha^{1/3}$, which solves to $\alpha \approx n^{1/4}$. In general, when $s \le \alpha^{3/2 - \delta}$ for some $\delta > 0$, it gives an $n^{1/3-\varepsilon}$-approximation.  To complete the proof of \Cref{thm:improved-sparsifier}, we develop a complementary argument which is stronger when $s$ is large. 

\paragraph{Density Increment Argument.}

A density increment argument is a method for locating an unusually
dense smaller subgraph inside a larger graph.  By zooming in to a much
smaller part of the graph while preserving, or increasing, the density, we obtain 
a relatively dense instance at a more useful scale.  Density increment arguments
are classical in additive combinatorics
\cite{roth1953certain,gowers2001new} and have also found important
applications in graph theory \cite{norin2023breaking}.

We use this idea to extract a smaller dense matching contractor from the original graph.  Starting from a matching contractor $G$
with $n$ left vertices whose matchings each have size about $s$, we identify a small subgraph $G_1$ that has only about $s$ left vertices but retains essentially the same edge density as the original graph. Moreover, $G_1$ is also a matching contractor. Our density increment lemma can be formalized as follows.

\begin{boxedtheorem}
\begin{lemma}[Density Increment Lemma]
\label{lem:DIL}
Let $G = (L\cup R,E)$ be an $\alpha$-matching-contractor composed of $k$ matchings
$M_1,\ldots,M_k$ each of size $s$. Then we can obtain a subgraph
$G_1 = (L_1\cup R_1,E_1)$ of $G$ such that:
\begin{enumerate}
    \item $|L_1|=\Theta(s)$,
    \item $|E_1|/|L_1|=\Omegaish( |E|/|L| )$, 
    \item $G_1$ can be decomposed into $k_1$ matchings $M_1^1, \dots, M_{k_1}^1$, each of size $|M_i^1| \le O(s/\alpha)$,  
    \item $G_1$ is $\alpha_1$-matching-contractor, where
          $\alpha_1\geq\Omegaish(\alpha^2s/n)$,
\end{enumerate}
\end{lemma}
\end{boxedtheorem}
\begin{proof}[Proof Ideas]
    Sample a matching $M_i$ uniformly at random from $\{M_i\}_{i \in [k]}$. Let $L_1 := L(M_i)$. Let $R_1 := N_{G-M_i}(L(M_i))$. We define graph $G_1$ to be $G_1 := G[L_1 \cup R_1]$. Note that $|L_1|=s$, since each matching $M_i$ has size $|M_i|=s$.     
    For each $j \in [k]$, let $M_j^1 := M_j \cap G_1$. Then graph $G_1$ is decomposed by the matchings $\{M_j^1\}_{j \in [k]}$, and each of these matchings has size at most $|M_j^1| \le s/\alpha$ by the Dispersion Lemma.
    Additionally, the density of $G_1$ is roughly the same as the density of $G$, since each vertex in $L_1$ has at most one edge not in $R_1$. 
    The only remaining property to show is that $G_1$ is an $\alpha_1$-matching-contractor. This follows from a straightforward double-counting argument.
    (See \Cref{lem:density} for details.)   
\end{proof}

This density increment lemma is most effective when $s$ is large because the contraction guarantee
\(
    \alpha_1\geq\Omegaish(\alpha^2s/n)
\)
becomes stronger as $s$ grows. Applying
Theorem~\ref{thm:intro-small-s} to $G_1$ yields
\(
    \alpha
    \leq
    \Oish\left(\frac{n^{3/7}}{s^{3/14}}\right).
\)
 Combining this
decreasing-with-$s$ bound with the growing-with-$s$ bound
\(
    \alpha\leq\Oish\left(n^{1/6}s^{1/3}\right)
\)
from Theorem~\ref{thm:intro-small-s}, the two bounds balance when
$s=n^{11/23}$, yielding
\[
    \alpha\leq\Oish\left(n^{15/46}\right).
\]
This completes our proof sketch of \Cref{thm:improved-sparsifier}.

\subsection{Streaming Algorithm for Load-Balancing Sparsifiers}
\label{subsec:stream-LBS}
Our streaming upper bound adapts the existential construction of Konrad and
Ros\'en~\cite{konrad2016approximating}, which gives an
$O(n^{1/3})$-load-balancing sparsifier with $O(n)$ edges. The basic object in
their construction is a \textit{semi-matching}.
\begin{boxeddefinition}
    \begin{definition}[Semi-Matching]
        \label{def:semi-matching}
        Let $G=(L\cup R,E)$ be an assignment graph. A semi-matching is a set of edges
        $S\subseteq E$ containing exactly one edge incident to every vertex in
        $L$.
        A semi-matching assigns each left vertex to a right vertex, and the load of
        $b\in R$ is $\deg_S(b)$.
    \end{definition}
\end{boxeddefinition}
Suppose that a semi-matching $S$ assigns vertex $a \in L$ to vertex $b \in R$, but $a$ is also adjacent to vertex $b' \in R$ in $G$. If
$\deg_S(b)\geq\deg_S(b')+2$, then
replacing edge $(a, b)$ with edge $(a, b')$ in $S$
 makes the loads in $S$ more balanced. We call $(b,a,b')$ a \emph{degree-minimizing path of length two}, and
call $S$ \emph{degree-minimizing} if it admits no such path. The semi-matchings
used by Konrad and Ros\'en are degree-minimizing in this sense.
\paragraph{The two-level construction of Konrad and Ros\'en.}
Their sparsifier stacks two levels of degree-minimizing semi-matchings.
\begin{enumerate}
    \item \emph{First level.} Fix a degree-minimizing semi-matching $S$ of $G$.
    For each $b\in R$, let
    \[
        L_b:=\{a\in L:(a,b)\in S\}
    \]
    be the set of left vertices that $S$ assigns to $b$. We call $L_b$ the
    \emph{group} of $b$. The groups partition $L$, and $|L_b|=\deg_S(b)$ is the
    load of $b$.
    \item \emph{Second level.} For each nonempty group $L_b$, fix a
    degree-minimizing semi-matching $S_b$ of the subgraph $G[L_b\cup R]$. Note
    that $S_b$ may assign the vertices of $L_b$ to right vertices other than $b$.
\end{enumerate}
The final load-balancing sparsifier of~\cite{konrad2016approximating} is $H_{\mathrm{KR}}:=S\cup\bigcup_{b\in R}S_b$. Each level has
$n:=|L|$ edges, so $H_{\mathrm{KR}}$ has at most $2n$ edges.
\paragraph{The charging inequality.}
The analysis of Konrad and Ros\'en rests on one key property of degree-minimizing
semi-matchings. Call a set $X\subseteq L$ \emph{matchable} in $G$ if some
matching of $G$ covers every vertex of $X$.
\begin{boxedtheorem}
    \begin{lemma}[Charging Inequality; cf.\ {\cite[Lemma~6]{konrad2016approximating}}]
        \label{lem:charging}
        Let $n=|L|$, and let $S$ be a degree-minimizing semi-matching of $G$
        with groups $L_b$ for $b \in R$. Every set $X\subseteq L$ that is matchable in $G$
        satisfies
        \[
            \sum_{b\in R}|X\cap L_b|\cdot |L_b|\leq 2n.
        \]
    \end{lemma}
\end{boxedtheorem}
We can interpret \cref{lem:charging} as follows. If we charge each vertex  $a \in X$ the size of the group it belongs to, i.e., the size $|L_b|$ where $a \in L_b$, then the total charge
is at most $2n$.
Thus a matchable set cannot contain many vertices from large groups.
\begin{proof}[Proof sketch]
Fix a matching $M$ of $G$ that covers $X$. Consider $a\in X$, let $b$ be the
vertex that $S$ assigns to $a$, so that $a$ is charged $|L_b|=\deg_S(b)$, and
let $b'$ be the partner of $a$ in matching $M$. Then $\deg_S(b)\leq\deg_S(b')+1$, since $(b,a,b')$ cannot be a degree-minimizing path of
length two. Sum this bound over all $a\in X$. The left side is the total charge, and
since the partners $b'$ in matching $M$ are distinct, the right side is at most
$\sum_{b\in R}\deg_S(b)+|X|=n+|X|\leq 2n$.
\end{proof}
\paragraph{Our streaming algorithm.}
It is not possible to directly compute Konrad and Ros\'en's load-balancing sparsifier in the streaming model: degree-minimizing semi-matchings are already an $O(\log n)$-approximation for load balancing~\cite{konrad2016approximating}, so a one-pass
$\Oish(n)$-space algorithm that computed one would contradict the
$n^{1/4-o(1)}$ lower bound of~\cite{assadi2025streaming}.

Instead of a semi-matching, our algorithm maintains a table $P$ with one
entry $P[a,i]$ for each left vertex $a\in L$ and each index
$i\in\{0,1,\dots,O(\log n)\}$. Every entry is either empty or holds one edge
incident to $a$, and we write $H$ for the subgraph formed by the currently
stored edges. When an edge $(a,b)$ arrives, the algorithm reads the current
degree of $b$ in $H$, computes the index
\[
    i:=\lfloor\log_2(\deg_H(b)+1)\rfloor,
    \qquad\text{i.e., } \quad 2^i\leq\deg_H(b)+1<2^{i+1},
\]
and stores edge $(a,b)$ in $P[a,i]$ if that entry is empty.  Otherwise the edge is
discarded, and we say that it is \emph{blocked}.
Thus, the buckets are maintained separately for each left vertex, while an edge’s bucket is determined by the degree of its right endpoint when it arrives. 
Each entry is filled at most
once, so $H$ has $O(n\log n)$ edges, and the algorithm makes a single pass. 

The key claim is that $H$ recovers the charging lemma (\cref{lem:charging}) up to a logarithmic
loss.  Specifically, the role of the groups $L_b$ in \cref{lem:charging} is now played by the set of \emph{buckets}
\[B_{b,i}:=\{a\in L:P[a,i]=(a,b)\}.\] Every edge $(a, b)$ associated with bucket $B_{b,i}$ was stored at a
time when $\deg_H(b)+1\in[2^i,2^{i+1})$, and storing it increased $\deg_H(b)$
by one, so $|B_{b,i}|\leq 2^i$. A blocked edge now plays the role of a missing
degree-minimizing path. Suppose that $(a,b')$ arrives, its index $i$ is
computed from $\deg_H(b')$, and it is blocked because $P[a,i]$ already holds
$(a,b)$. Then $a$ lies in the bucket $B_{b,i}$, and
\[
    |B_{b,i}|\leq 2^i\leq\deg_H(b')+1.
\]
This is identical to the claim $\deg_S(b)\leq\deg_S(b')+1$ from the proof of
\cref{lem:charging}, with the group $L_b$ replaced by the bucket $B_{b,i}$
and loads in $S$ replaced by degrees in $H$. Summing it over a matchable set
as before gives the charging inequality for buckets, with $2n$ replaced by
$|E(H)|+n=O(n\log n)$.
This charging inequality for $H$ is formalized in  \cref{lem:streaming-first-charge}  of \cref{sec:streaming-sparsifiers}.

Now, just like in Konrad and Ros\'en's two-level construction, we can run our algorithm a second time inside every bucket $B_{b,i}$, with a separate table that processes each member's edges from the moment it joins the bucket. This gives a second-level charging
inequality inside each bucket (\cref{lem:streaming-second-charge}). 
Across both levels, the algorithm uses $O(n\log^2 n)$ words of space and
stores $O(n\log^2 n)$ edges. The analysis of $H_{\mathrm{KR}}$ in~\cite{konrad2016approximating} only uses the charging inequality guaranteed by the semi-matchings $S$ and $S_b$.
Thus, by a similar  argument, we can show that the stored edges in $H$ form an 
$\Oish(n^{1/3})$-load-balancing sparsifier. This completes the proof sketch of 
\cref{thm:streaming-sparsifier}.

\subsection{Load-Balancing Sparsifier Lower Bound }
\label{subsec:lower}

We prove our lower bound for load-balancing sparsifiers by constructing dense matching contractors. By the Matching Contractor Duality in \Cref{lem:MC_duality}, such a construction translates directly into the sparsifier lower bound in \Cref{thm:linear-size-lower}.

Our starting point is the general construction underlying Theorem~3 of~\cite{assadi2025streaming}. Their construction adds edges to the matching contractor in rounds. Each round $i$ is governed by a projection determined by a set $S_i$, and the edges added in that round decompose into matchings. The pairwise intersections among the sets $\{S_i\}_i$ control the residual neighborhoods of these matchings, and hence the contraction parameter of the resulting graph. The authors of~\cite{assadi2025streaming} obtain a suitable set family $\{S_i\}_i$ from the classical Gilbert--Varshamov bound in coding theory.

We follow the same round-based framework, but use a different coding-theoretic ingredient. We replace the Gilbert--Varshamov set family with an explicit equidistant binary code meeting the Plotkin bound. This gives $t$ sign vectors such that every pair disagrees on a $1/2+\Omega(1/t)$ fraction of coordinates. Each sign vector defines a round of the graph construction through a simple linear map over a finite field. For a matching created in one round, its neighbors in another round have fixed values in every coordinate where the two sign vectors disagree. Thus, the large pairwise distances between the sign vectors keep the residual neighborhoods small. The resulting matching contractor has $\Theta(tn)$ edges and contraction parameter $\Omega(n^{1/4+1/(4t)})$. Taking $t=\Theta(c)$ and applying \Cref{lem:MC_duality} proves \Cref{thm:linear-size-lower}.

\section{Preliminaries}

Throughout this paper, for a bipartite graph $G=(L\cup R,E)$, we write $n=|L|$ for the number of clients on the left side of $G$.

Here we formally define the computational models we use in this paper. In these models, we assume that vertex identifiers use $O(\log n)$ bits. We begin with the two-party one-way communication model for graph problems. 

\begin{boxeddefinition}
\begin{definition}[Two-party one-way communication model]
\label{def:comm}
Let \(G=(L\cup R,E)\) be a bipartite graph whose edges are partitioned into disjoint sets
\(E_A\) and \(E_B\). Alice receives \(G_A=(L\cup R,E_A)\), Bob receives
\(G_B=(L\cup R,E_B)\), and the vertex set is known to both players. Alice sends
one message to Bob, after which Bob outputs a solution to the graph problem
on \(G=(L\cup R,E_A\cup E_B)\). The communication cost is the maximum length $s(n)$ of
Alice's message, in bits.
\end{definition}
\end{boxeddefinition}

Likewise, we define the streaming model for graph algorithms.

\begin{boxeddefinition}
\begin{definition}[Graph streaming model]
\label{def:streaming}
Let \(G=(L\cup R,E)\) be a simple bipartite graph whose vertex set is known in advance. In a graph stream, the
edges of \(E\) arrive one at a time in an arbitrary order, with each edge appearing exactly once. A \(p(n)\)-pass  streaming algorithm scans a graph stream  \(p\) times, and outputs a solution after its final pass.  Its space usage is the maximum amount of working memory $s(n)$ needed at any time, in bits.
\end{definition}
\end{boxeddefinition}

\section{Load-Balancing Sparsifier Upper Bound}
\label{sec:upper}

In this section, we prove our load-balancing sparsifier upper bound, which we
restate below.

\begin{boxedtheorem}
    \lbsImprovedSparsifier*
\end{boxedtheorem}

By the duality between load-balancing sparsifiers and matching contractors in
\cref{lem:MC_duality}, to prove this theorem it is enough to show that, for
some absolute constant \(c>0\), every
\(c n^{1/3-\varepsilon_0}\)-matching-contractor has \(O(n\log^2 n)\)
edges.  The additional \(O(\log n)\) factor comes from the duality lemma.

We prove this statement by contrapositive.  Recall that
\[
\begin{aligned}
    \textsf{Contractor}(n,\alpha)
    &:=\max\{|E(G)|:G\text{ is an }\alpha\text{-MC with }n\text{ left vertices}\},\\
    \textsf{d}(n,\alpha)&:=\frac{\textsf{Contractor}(n,\alpha)}{n}.
\end{aligned}
\]
Suppose that \(\textsf{d}(n,\alpha)>C_0\log^2 n\), where \(C_0>1\) is a
sufficiently large constant.  We will show that
\(\alpha<n^{1/3-\varepsilon_0}\).  Since \(\textsf{d}(n,\alpha)\) is
non-increasing in \(\alpha\), we may assume that \(\textsf{d}(n,\alpha) = \Theta(C_0\log^2 n)\).
In other words, we may fix an
\(\alpha\)-matching-contractor \(G\) with \(n\) left vertices and
\(\Theta(C_0n\log^2 n)\) edges.
We have therefore reduced the proof to the following statement.

\begin{theorem}[Matching-contractor upper bound]
\label{thm:mc-upper}
    Let \(C_0>1\) be a sufficiently large absolute constant.  There is an
    absolute constant \(\varepsilon_0>0\) such that, for all sufficiently
    large \(n\), every \(\alpha\)-matching-contractor \(G\) with
    \(n\) left vertices and \(|E(G)|=\Theta(C_0n\log^2 n)\) satisfies
    \(\alpha<n^{1/3-\varepsilon_0}\).
\end{theorem}

We use the following notation throughout the proof.

\begin{center}
\begin{tcolorbox}[
    enhanced,
    width=0.88\textwidth,
    colback=pastelblue,
    colframe=blue!40!black,
    boxrule=0.6pt,
    arc=3mm,
    left=4mm,
    right=4mm,
    top=2mm,
    bottom=2mm
]
\begin{itemize}
    \item \(\textsf{Contractor}(n,\alpha)\) is the maximum number of edges
    in an \(\alpha\)-matching-contractor with \(n\) left vertices, and
    \(\textsf{d}(n,\alpha):=\textsf{Contractor}(n,\alpha)/n\).
    \item Let \(G=(L\cup R,E)\) be an \(\alpha\)-matching-contractor with
    \(n=|L|\) left vertices, decomposed into matchings
    \(M_1,\ldots,M_k\).  Let
    \(\mathsf d_G:=|E|/n\) and
    \(s:=\frac1k\sum_{i\in[k]}|M_i|\).
    \item For each \(i\in[k]\), define
    \[
        L_i:=V(M_i)\cap L,\qquad
        R_i:=N_{G-M_i}(L_i),\qquad
        H_i:=(G-M_i)[L_i\cup R_i].
    \]
\end{itemize}
\end{tcolorbox}
\end{center}

\begin{lemma}[Normalization Lemma]
\label{lem:clean}
    Let \(F\) be a \(\beta\)-matching-contractor with \(N\) left vertices and density
    \(\mathsf d_F:=|E(F)|/N\ge C\log N\), where \(C\) is a sufficiently large absolute
    constant.  Then there is an \(\alpha\)-matching-contractor
    \(G=(L\cup R,E)\) such that:
    \begin{enumerate}
        \item \(G\) has \(n=\Theta(N)\) left vertices,
        \(\alpha=\Omega(\beta)\), and
        \(\mathsf d_G=\Theta(\mathsf d_F/\log N)\);
        \item every \(v\in L\) has
        \(\deg_G(v)=\Theta(\mathsf d_G)\);
        \item every \(M_i\) has size \(|M_i|=\Theta(s)\);
        \item if \(r\in R_i\), then
        \(\deg_{H_i}(r)=\Omega(\alpha)\).
        \item if \(F\) is extremal, then
        \(\mathsf d_G=\Omega(\textsf{d}(N,\beta)/\log N)\).
    \end{enumerate}
\end{lemma}
\begin{proof}
    We defer the proof to \cref{app:clean}.
\end{proof}

Apply \cref{lem:clean} and continue to call the resulting graph \(G\).  This
changes \(n\) and \(\alpha\) only by constant factors, and gives
\(\mathsf d_G=\Theta(C_0\log n)\) together with the other conclusions of the
lemma.

\begin{remark}
    The graphs \(H_i\) and the residual-degree guarantee in
    \cref{lem:clean} may seem mysterious now.  They will be useful when we
    study the neighborhoods of a fixed right vertex.
\end{remark}

We first give two warm-ups that introduce the main counting ideas.

\subsection{Warm-up \(\#1\)}
\label{sec:warmup1}

We begin with the simple bound \(\alpha=O(\sqrt n)\).  The main observation is
that any two sets \(L_i\) and \(L_j\) have small intersection.

\begin{lemma}[Dispersion lemma]
\label{lemma:intersect}
    For any two distinct indices \(i,j\in[k]\),
    \(|L_i\cap L_j|\le O(s/\alpha)\).
\end{lemma}
\begin{proof}
    Let \(M:=\{(u,v)\in M_j:u\in L_i\cap L_j\}\).
    Since \(i\ne j\), the right endpoints of \(M\) belong to \(R_i\).  Hence
    \[
        |L_i\cap L_j|=|M|=|V(M)\cap R|\le |R_i|\le |M_i|/\alpha
        =O(s/\alpha),
    \]
    since \(|M_i|=\Theta(s)\).
\end{proof}

\begin{remark}
    In the lower-bound construction of \cite{assadi2025streaming}, whenever
    \(L_i\cap L_j\ne\emptyset\), one has
    \(|L_i\cap L_j|=\Theta(s/\alpha)\).  Thus
    \cref{lemma:intersect} is tight in general.
\end{remark}

The pairwise overlap bound already forces \(G\) to contain many matchings.

\begin{claim}
\label{clm:large_k}
    \(k\ge\Omega(\alpha\mathsf d_G)\).
\end{claim}
\begin{proof}
    Fix \(L_1\).  Double counting gives
    \[
        \sum_{i=2}^k|L_1\cap L_i|
        =\sum_{v\in L_1}\deg_{G-M_1}(v)=|E(H_1)|
        =\Theta(s\mathsf d_G).
    \]
    Each summand is \(O(s/\alpha)\) by \cref{lemma:intersect}, so
    \(k\ge\Omega(\alpha\mathsf d_G)\).
\end{proof}

We can now combine this lower bound on \(k\) with the total edge count.

\begin{theorem}[First warm-up]
\label{thm:warmup1}
    \(\alpha=O(n^{1/2})\).
\end{theorem}
\begin{proof}
    Since \(\mathsf d_G\) is sufficiently large, each \(R_i\) is
    nonempty.  The matching-contractor condition therefore gives
    \(|M_i|=\Omega(\alpha)\), and hence \(s=\Omega(\alpha)\).  Thus
    \[
        \alpha^2\mathsf d_G\le O(s\alpha\mathsf d_G)
        \le O(ks)=O(n\mathsf d_G),
    \]
    where the middle inequality is \cref{clm:large_k} and
    \(ks=\Theta(n\mathsf d_G)\).  Cancelling \(\mathsf d_G\) proves the
    theorem.
\end{proof}

\subsection{Warm-up \(\#2\)}
\label{sec:warmup2}

For the second warm-up, we fix a right vertex \(r\) and look inside the graphs
\(H_i\).  This gives \(\alpha=O((n\mathsf d_G)^{1/3})=
O(n^{1/3}\log^{1/3}n)\), nearly recovering the previous upper bound of
\cite{konrad2016approximating} in \cref{thm:prior-upper}.

Recall that the graphs \(H_i\) were defined in \cref{sec:upper}.  For
\(r\in R\) and \(i\in[k]\), let
\[
    L_i^r:=N_{H_i}(r)\subseteq L_i,
    \qquad
    \mathcal L^r:=(L_i^r)_{i\in[k]:L_i^r\ne\emptyset}.
\]

\begin{claim}
\label{clm:lower}
    If \(L_i^r\ne\emptyset\), then \(|L_i^r|\ge\Omega(\alpha)\).
\end{claim}
\begin{proof}
    If \(L_i^r\ne\emptyset\), then \(r\in R_i\), so the last property of
    \cref{lem:clean} gives \(|L_i^r|=\deg_{H_i}(r)=\Omega(\alpha)\).
\end{proof}

This immediately gives an upper bound on the number of sets in
\(\mathcal L^r\).

\begin{claim}
\label{clm:warmup2_upper}
    For every \(r\in R\),
    \(|\mathcal L^r|\le O(k\mathsf d_G/\alpha)\).
\end{claim}
\begin{proof}
    Double counting gives
    \[
        \sum_{L_i^r\in\mathcal L^r}|L_i^r|
        =\sum_{v\in N_G(r)}(\deg_G(v)-1)
        =O(|N_G(r)|\mathsf d_G)=O(k\mathsf d_G).
    \]
    Here \(|N_G(r)|\le k\), since each matching contains at most one edge at
    \(r\).  The claim follows from \cref{clm:lower}.
\end{proof}

We also have a lower bound whenever \(\mathcal L^r\ne\emptyset\).

\begin{claim}
\label{clm:warmup2_lower}
    If \(\mathcal L^r\ne\emptyset\), then
    \(|\mathcal L^r|\ge\Omega(\alpha^2\mathsf d_G/s)\).
\end{claim}
\begin{proof}
    Fix \(L_x^r\in\mathcal L^r\).  Then
    \[
        \sum_{\substack{L_i^r\in\mathcal L^r\\i\ne x}}
        |L_x^r\cap L_i^r|
        =\sum_{v\in L_x^r}(\deg_G(v)-2)
        =\Omega(|L_x^r|\mathsf d_G)=\Omega(\alpha\mathsf d_G),
    \]
    by \cref{clm:lower}.  Each summand is at most
    \(|L_x\cap L_i|=O(s/\alpha)\) by \cref{lemma:intersect}, proving the
    claim.
\end{proof}

We are now ready to prove the upper bound on \(\alpha\).

\begin{theorem}[Second warm-up]
\label{thm:warmup2}
    \(\alpha\le O((n\mathsf d_G)^{1/3})\).  In particular,
    \(\alpha\le O(n^{1/3}\log^{1/3}n)\).
\end{theorem}
\begin{proof}
    Comparing \cref{clm:warmup2_upper} with \cref{clm:warmup2_lower} gives
    \[
        \frac{\alpha^2\mathsf d_G}{s}
        \le O\left(\frac{k\mathsf d_G}{\alpha}\right)
        \quad\Longrightarrow\quad
        \alpha^3\le O(ks)=O(n\mathsf d_G),
    \]
    which is the claimed bound.
\end{proof}

\subsection{A Stronger Dispersion Argument}

The preceding warm-up counts intersections with one fixed set \(L_x^r\).  By
counting many of these intersections together, we obtain a stronger bound.

\begin{theorem}
\label{thm:final_dispersion}
    \(\alpha\le O(n^{1/6}s^{1/3}\mathsf d_G^{-1/6})\).
\end{theorem}

Since \(\mathsf d_G=\Theta(\log n)\), this is
\(\alpha\le\Oish(n^{1/6}s^{1/3}\mathsf d_G^{-1/6})\).

\begin{remark}
    When \(s=\Theta(\alpha)\), \cref{thm:final_dispersion} gives
    \(\alpha=O(n^{1/4}\mathsf d_G^{-1/4})=\Oish(n^{1/4})\).
\end{remark}

\begin{restatable}[Stronger dispersion lemma]{lemma}{strongerdispersionlemma}
\label{lem:final_dispersion}
    There exist \(r\in R\), a subfamily
    \(\mathcal X\subseteq\mathcal L^r\), and a set \(U\subseteq L\) such
    that:
    \begin{enumerate}
        \item
        \[
            \sum_{L_i^r\in\mathcal L^r\setminus\mathcal X}
            |L_i^r\cap U|
            \ge
            \Omega\left(\frac{\alpha^3\mathsf d_G^3}{s}\right);
        \]
        \item for every \(L_i^r\in\mathcal L^r\setminus\mathcal X\),
        \(|L_i^r\cap U|\le O(s^2/\alpha^2)\).
    \end{enumerate}
\end{restatable}

Before proving the lemma, let us see how it gives
\cref{thm:final_dispersion}.

\begin{claim}
\label{clm:final_dispersion}
    For the vertex \(r\) in \cref{lem:final_dispersion},
    \(|\mathcal L^r|\ge\Omega(\alpha^5\mathsf d_G^3/s^3)\).
\end{claim}
\begin{proof}
    By the two conclusions of \cref{lem:final_dispersion},
    \[
    \begin{aligned}
        |\mathcal L^r|
        &\ge
        \left|
            \{L_i^r\in\mathcal L^r\setminus\mathcal X:
              L_i^r\cap U\ne\emptyset\}
        \right|\\
        &\ge
        \frac{
            \sum_{L_i^r\in\mathcal L^r\setminus\mathcal X}
            |L_i^r\cap U|
        }{
            \max_{L_i^r\in\mathcal L^r\setminus\mathcal X}
            |L_i^r\cap U|
        }\\
        &\ge
        \Omega\left(
            \frac{\alpha^3\mathsf d_G^3/s}{s^2/\alpha^2}
        \right)
        =
        \Omega\left(\frac{\alpha^5\mathsf d_G^3}{s^3}\right).
    \end{aligned}
    \]
\end{proof}

Assuming \cref{lem:final_dispersion}, we can now finish the proof of the theorem.

\begin{proof}[Proof of \cref{thm:final_dispersion}]
    Combining \cref{clm:warmup2_upper} with \cref{clm:final_dispersion},
    and using \(ks=\Theta(n\mathsf d_G)\), we obtain
    \[
        \frac{\alpha^5\mathsf d_G^3}{s^3}
        \le O\left(\frac{k\mathsf d_G}{\alpha}\right)
        \quad\Longrightarrow\quad
        \frac{\alpha^6}{s^2}\le O\left(\frac{n}{\mathsf d_G}\right),
    \]
    and the stated bound follows.
\end{proof}

\subsubsection{Proof of the Stronger Dispersion Lemma}

We first prove two counting claims.  The first lower-bounds the total overlap,
and the second uses averaging to choose a useful pair \((i,r)\).

\begin{claim}
\label{clm:pair_square_mass}
    \[
        \sum_{i\in[k]}
        \sum_{\substack{u,v\in R\\u\ne v}}
        |L_i^u\cap L_i^v|^2
        \ge
        \Omega(k\alpha^2\mathsf d_G^4).
    \]
\end{claim}
\begin{proof}
    Fix \(i\in[k]\), and set
    \(A_i:=\sum_{\substack{u,v\in R\\u\ne v}}|L_i^u\cap L_i^v|\).
    Counting triples \((x,u,v)\) with \(x\in L_i\), \(u\ne v\), and both
    \(u\) and \(v\) adjacent to \(x\) in \(H_i\), we get
    \[
        A_i
        =
        \sum_{x\in L_i}
        \deg_{H_i}(x)\bigl(\deg_{H_i}(x)-1\bigr).
    \]
    The graph \(H_i\) keeps every edge at \(x\in L_i\) except its edge in
    \(M_i\), so \(\deg_{H_i}(x)=\Theta(\mathsf d_G)\).  Since
    \(|L_i|=\Theta(s)\) and \(\mathsf d_G\) is sufficiently large, we have
    \(A_i=\Omega(s\mathsf d_G^2)\).  Moreover, only pairs in \(R_i\) contribute
    and \(|R_i|=O(s/\alpha)\).  Cauchy--Schwarz therefore gives
    \[
    \begin{aligned}
        \sum_{\substack{u,v\in R\\u\ne v}}
        |L_i^u\cap L_i^v|^2
        &\ge
        \frac{A_i^2}{|R_i|^2}\\
        &\ge
        \frac{\Omega(s^2\mathsf d_G^4)}
             {O(s^2/\alpha^2)}
        =
        \Omega(\alpha^2\mathsf d_G^4).
    \end{aligned}
    \]
    Summing over \(i\in[k]\) proves the claim.
\end{proof}

\begin{definition}[Hitting a pair]
    We say that \(M_i\) hits \((i',v)\in[k]\times R\) if \(i'\ne i\) and
    \((L_{i'}^v\times\{v\})\cap E(M_i)\ne\emptyset\).  Let \(Q_i\) be the
    set of pairs hit by \(M_i\).
\end{definition}

\begin{claim}
\label{clm:hitting_pair}
    There is a pair \((i,r)\in[k]\times R\) such that
    \[
        \sum_{(i',v)\in Q_i}
        |L_{i'}^v\cap L_{i'}^r|
        \ge
        \Omega\left(\frac{\alpha^3\mathsf d_G^4}{s}\right).
    \]
\end{claim}
\begin{proof}
    Let
    \[
    \begin{aligned}
        \mathcal P&:=\{(i,r)\in[k]\times R:r\in R_i\},\\
        F(i,r)&:=\sum_{(i',v)\in Q_i}|L_{i'}^v\cap L_{i'}^r|.
    \end{aligned}
    \]
    We first show that
    \[
        \sum_{(i,r)\in\mathcal P}F(i,r)
        \ge
        \Omega(k\alpha^2\mathsf d_G^4).
    \]

    To see this, fix \(i'\in[k]\), distinct \(u,v\in R\), and
    \(x\in L_{i'}^u\cap L_{i'}^v\).  The edge \((x,v)\) belongs to some
    matching \(M_j\) with \(j\ne i'\).  Because \(M_j\) is a matching and
    \(u\ne v\), the edge \((x,u)\) does not belong to \(M_j\).  Thus
    \(x\in L_j\), \(u\in R_j\), and \((i',v)\in Q_j\).  As \(x\)
    ranges over \(L_{i'}^u\cap L_{i'}^v\), the indices \(j\) are distinct:
    all corresponding edges \((x,v)\) share the endpoint \(v\), and a
    matching contains at most one such edge.  The fixed triple
    \((i',u,v)\) therefore contributes at least
    \(|L_{i'}^u\cap L_{i'}^v|^2\) to the displayed sum.  Summing and using
    \cref{clm:pair_square_mass} proves this bound.

    Finally, \(|\mathcal P|=\sum_i|R_i|=O(ks/\alpha)\).  Averaging over
    \(\mathcal P\) completes the proof.
\end{proof}

We can now prove the lemma.

\strongerdispersionlemma*
\begin{proof}
    Let \((i,r)\) be the pair from \cref{clm:hitting_pair}, put \(Q:=Q_i\),
    and, for \(q=(i_q,v_q)\in Q\), let
    \(A_q:=L_{i_q}^{v_q}\cap L_{i_q}^r\).
    Then
    \[
        \sum_{q\in Q}|A_q|
        \ge
        \Omega\left(\frac{\alpha^3\mathsf d_G^4}{s}\right).
    \]

    For each vertex \(v\) appearing in \(Q\), retain a pair
    \((j,v)\in Q\) maximizing \(|A_{(j,v)}|\), and let \(Q'\) be the set of
    retained pairs.
    For a fixed \(v\), all pairs \((j,v)\in Q\) are witnessed by the same
    edge \((x,v)\in M_i\).  Every such \(j\) satisfies \(x\in L_j\), so
    there are at most \(\deg_G(x)=O(\mathsf d_G)\) choices.  Keeping the
    largest \(A_{(j,v)}\) therefore loses at most this factor, and hence
    \[
        W:=\sum_{q\in Q'}|A_q|
        \ge
        \Omega\left(\frac{\alpha^3\mathsf d_G^3}{s}\right).
    \]
    Let \(U:=\bigcup_{q\in Q'}A_q\).

    We next identify the sets whose intersection with \(U\) is unusually
    large.  Let \(C\) be a sufficiently large absolute constant and define
    \[
        \mathcal X
        :=
        \left\{
            L_j^r\in\mathcal L^r:
            |L_j^r\cap U|>C\frac{s^2}{\alpha^2}
        \right\}.
    \]
    Then item 2 holds by definition.

    For \(q=(i_q,v_q)\in Q'\) and \(x\in A_q\), call \(j\in[k]\) good for
    \((q,x)\) if \(x\in L_j^r\), \(j\ne i_q\), and
    \((x,v_q)\notin E(M_j)\).
    The number of indices \(j\) with \(x\in L_j^r\) is
    \(\deg_G(x)-1=\Theta(\mathsf d_G)\).  Excluding \(i_q\) and the
    matching containing \((x,v_q)\) leaves
    \(\Omega(\mathsf d_G)\) good choices.  Thus
    \[
        \mathcal T:=
        \{(q,x,j):q\in Q',\ x\in A_q,\ j\text{ is good for }(q,x)\}.
    \]
    We have \(|\mathcal T|\ge\Omega(\mathsf d_GW)\).

    Let
    \[
        \mathcal I_{\rm good}
        :=
        \{(j,x)\in[k]\times U:
          (q,x,j)\in\mathcal T\text{ for some }q\in Q'\}.
    \]
    A fixed incidence \((j,x)\) arises from at most
    \(O(\mathsf d_G)\) triples in \(\mathcal T\): the vertices \(v_q\)
    are distinct and each is adjacent to \(x\).  Consequently,
    \(|\mathcal I_{\rm good}|\ge\Omega(W)\).

    We also need a per-set bound.  Fix \(j\in[k]\).  If
    \((j,x)\in\mathcal I_{\rm good}\) is witnessed by
    \(q=(i_q,v_q)\), then \(v_q\in R_j\).  The vertices \(v_q\) are
    distinct, so there are at most \(|R_j|=O(s/\alpha)\) possible witnesses
    \(q\).  For each one, \(|L_j^r\cap A_q|\le |L_j\cap L_{i_q}|=
    O(s/\alpha)\) by \cref{lemma:intersect}.  Thus each \(j\) has at most
    \(O(s^2/\alpha^2)\) good incidences.

    Let
    \[
        \mathcal I
        :=\{(j,x)\in[k]\times U:x\in L_j^r\},
        \qquad
        \mathcal I_{\rm bad}
        :=\mathcal I\setminus\mathcal I_{\rm good}.
    \]
    For a bad incidence \((j,x)\), choose \(q=(i_q,v_q)\in Q'\) with
    \(x\in A_q\).
    Since \((j,x)\) is not good, either \(j=i_q\) or
    \((x,v_q)\in E(M_j)\).  For a fixed occurrence \((q,x)\), at most
    two indices \(j\) satisfy these alternatives.  Charging \((j,x)\) to
    \((q,x)\) gives \(|\mathcal I_{\rm bad}|\le2\sum_{q\in Q'}|A_q|=2W\).

    Choose an absolute constant \(C_{\mathrm{inc}}\) so that every \(j\) has
    at most \(C_{\mathrm{inc}}s^2/\alpha^2\) good incidences.  If
    \(L_j^r\in\mathcal X\), then it has at least
    \((C-C_{\mathrm{inc}})s^2/\alpha^2\) bad incidences.  Thus the sets in
    \(\mathcal X\) contain at most
    \(\frac{C_{\mathrm{inc}}}{C-C_{\mathrm{inc}}}|\mathcal I_{\rm bad}|
    =O(W/C)\) good incidences.
    Taking \(C\) large enough leaves \(\Omega(W)\) good incidences in
    sets outside \(\mathcal X\).  Hence
    \[
        \sum_{L_j^r\in\mathcal L^r\setminus\mathcal X}
        |L_j^r\cap U|
        \ge
        \Omega(W)
        \ge
        \Omega\left(\frac{\alpha^3\mathsf d_G^3}{s}\right).
    \]
    This proves item 1.
\end{proof}

\subsection{Density Increment Lemma}
\label{subsec:density-increment}

The stronger dispersion bound is most useful when the matchings \(M_i\) are
small.  For large \(s\), the next lemma extracts a smaller matching-contractor
of essentially the same density.  See \cref{subsec:LBS-over} for an overview
of the density-increment argument.

\begin{lemma}[Density increment lemma]
\label{lem:density}
    Let \(G=(L\cup R,E)\) be the normalized
    \(\alpha\)-matching-contractor above, decomposed into \(k\) matchings,
    each of size \(\Theta(s)\).  There is a subgraph
    \(G_1=(V_1,E_1)\subseteq G\) such that:
    \begin{enumerate}
        \item \(n_1:=|V_1\cap L|=\Theta(s)\);
        \item \(\mathsf d_{G_1}:=|E_1|/n_1=\Theta(\mathsf d_G)\);
        \item \(G_1\) is an \(\alpha_1\)-matching-contractor, where
        \(\alpha_1\ge\Omega(\alpha^2s/(n\mathsf d_G))\);
        \item \(G_1\) is decomposed into matchings
        \(M_1^1,\ldots,M_{k_1}^1\), each of size \(O(s/\alpha)\).
    \end{enumerate}
\end{lemma}

\begin{proof}
    Let \(H_1:=(G-M_1)[L_1\cup R_1]\).  We have
    \(|L_1|=\Theta(s)\) and \(|R_1|\le |M_1|/\alpha=O(s/\alpha)\).
    Moreover,
    \[
        |E(H_1)|=\sum_{v\in L_1}\bigl(\deg_G(v)-1\bigr)
        =\Theta(s\mathsf d_G),
    \]
    since the left degrees are \(\Theta(\mathsf d_G)\) and
    \(\mathsf d_G\) is sufficiently large.

    For each \(i\ge2\), let \(M_i^1:=M_i\cap E(H_1)\).
    These matchings partition \(E(H_1)\).  Since every right endpoint of
    \(M_i^1\) lies in \(R_1\), we have \(|M_i^1|=O(s/\alpha)\).  Let \(J\)
    be the set of indices for which \(M_i^1\ne\emptyset\).  For \(i\in J\),
    put
    \[
        L_i^1:=V(M_i^1)\cap L=L_i\cap L_1,
        \qquad
        R_i^1:=N_{H_1-M_i^1}(L_i^1).
    \]
    Put \(A:=\sum_{i\in J}|L_i^1|\) and
    \(B:=\sum_{i\in J}|R_i^1|\).  The matching partition gives
    \(A=|E(H_1)|=\Theta(s\mathsf d_G)\).
    Also \(R_i^1\subseteq R_i\cap R_1\), and therefore
    \[
    \begin{aligned}
        B
        &\le
        \sum_{i=2}^k|R_i\cap R_1|\\
        &=
        \sum_{r\in R_1}(|\mathcal L^r|-1)\\
        &\le
        |R_1|\cdot O\left(\frac{k\mathsf d_G}{\alpha}\right)\\
        &\le
        O\left(\frac{ks\mathsf d_G}{\alpha^2}\right),
    \end{aligned}
    \]
    where the third line uses \cref{clm:warmup2_upper}.

    If \(B=0\), the result follows with \(G_1=H_1\).  Otherwise, set
    \[
        \tau:=\frac{A}{2B},
        \qquad
        I:=\{i\in J:|L_i^1|\ge\tau|R_i^1|\}.
    \]
    The discarded indices account for less than half the edges, because
    \[
        \sum_{i\in J\setminus I}|L_i^1|
        <
        \tau\sum_{i\in J\setminus I}|R_i^1|
        \le \tau B
        =A/2.
    \]
    Let \(G_1\) be the union of the matchings indexed by \(I\), on the vertex
    set \(V(H_1)\).  It retains at least \(A/2\) edges, so
    \(|E(G_1)|/|L_1|=\Theta(\mathsf d_G)\).
    Deleting the other matchings can only shrink residual neighborhoods.
    Hence every retained matching satisfies
    \[
        \bigl|
        N_{G_1-M_i^1}(L_i^1)
        \bigr|
        \le |R_i^1|
        \le |M_i^1|/\tau,
    \]
    and \(G_1\) is a \(\tau\)-matching-contractor.  Finally,
    \(\tau=\Omega(\alpha^2/k)=\Omega(\alpha^2s/(n\mathsf d_G))\), since
    \(ks=\Theta(n\mathsf d_G)\).  This proves the lemma.
\end{proof}

Before using \(G_1\), we apply \cref{lem:clean} once more.  Since
\(\mathsf d_{G_1}=\Theta(\mathsf d_G)=\Theta(\log n)\) and
\(n_1=|L_1|=\Theta(s)\le n\), the lemma applies when \(C_0\) is sufficiently
large.  It gives a normalized graph \(\widehat G_1\) whose parameters satisfy
\[
\begin{aligned}
    \widehat n_1&=\Theta(s), &
    \widehat{\mathsf d}_1
        &\ge\Omega\left(\frac{\mathsf d_G}{\log n}\right),\\
    \widehat\alpha_1
        &\ge\Omega\left(\frac{\alpha^2s}{n\mathsf d_G}\right), &
    \widehat s_1&\le O(s/\alpha).
\end{aligned}
\]
The last bound holds because normalization does not increase matching sizes.
This is the second and final logarithmic loss in the matching-contractor
argument.

\subsection{Completing the Upper Bound}

We finish by balancing the direct dispersion bound with the bound obtained
after the density increment.

\begin{proof}[Proof of \cref{thm:mc-upper}]
    Write \(d:=\mathsf d_G=\Theta(\log n)\).  By
    \cref{thm:final_dispersion},
    \begin{equation}
        \alpha
        \le
        O\left(n^{1/6}s^{1/3}d^{-1/6}\right).
        \label{eq:direct-dispersion-bound}
    \end{equation}

    Applying \cref{thm:final_dispersion} to the normalized graph
    \(\widehat G_1\) above gives
    \[
        \frac{\widehat\alpha_1^6}{\widehat s_1^2}
        \le
        O\left(\frac{\widehat n_1}{\widehat{\mathsf d}_1}\right)
        \le
        O\left(\frac{s\log n}{d}\right).
    \]
    Since \(\widehat\alpha_1=\Omega(\alpha^2s/(nd))\) and
    \(\widehat s_1=O(s/\alpha)\), substitution gives
    \(\alpha^{14}s^3\le O(n^6d^5\log n)\), and hence
    \begin{equation}
        \alpha
        \le
        O\left(
            n^{3/7}
            d^{5/14}
            (\log n)^{1/14}
            s^{-3/14}
        \right).
        \label{eq:density-increment-bound}
    \end{equation}

    The right side of \cref{eq:direct-dispersion-bound} grows with \(s\),
    while the right side of \cref{eq:density-increment-bound} decreases
    with \(s\).  They are equal when
    \(s=n^{11/23}d^{22/23}(\log n)^{3/23}\).  Thus
    \[
        \alpha
        \le O\left(n^{15/46}d^{7/46}(\log n)^{1/23}\right)
        =O\left(n^{15/46}\log^{9/46}n\right),
    \]
    where we used \(d=\Theta(\log n)\).  Since
    \(15/46=1/3-1/138\), we may take
    \(\varepsilon_0=0.007\).  This proves
    \cref{thm:mc-upper}: there is an
    absolute constant \(c_0>0\) such that
    \[
        \textsf{Contractor}
        \bigl(n,c_0 n^{1/3-\varepsilon_0}\bigr)
        =
        O(n\log^2 n).
    \]
\end{proof}

We can now prove \cref{thm:improved-sparsifier}.

\lbsImprovedSparsifier*
\begin{proof}[Proof of \cref{thm:improved-sparsifier}]
    Set \(\beta:=c_0 n^{1/3-\varepsilon_0}\), where \(c_0\) and
    \(\varepsilon_0\) are given above.  The upper-bound
    direction of \cref{lem:MC_duality} gives
    \[
    \begin{aligned}
        \textsf{Sparsifier}(n,2\beta)
        &\le
        O(\log n)\,
        \textsf{Contractor}(n,\beta)\\
        &=
        O(n\log^3 n).
    \end{aligned}
    \]
\end{proof}

We can also complete the proof of \cref{cor:communication}.
We use the following proposition from~\cite{konrad2016approximating,assadi2025streaming}.

\begin{proposition}
\label{prop:lbs-communication}
Fix \(n,\alpha\ge1\), and suppose every \(n\)-client assignment graph admits an
\(\alpha\)-load-balancing sparsifier with at most \(m\) edges. Then there is a deterministic
one-way communication protocol with cost \(O(m\log n)\) bits that computes an
\((\alpha+1)\)-approximation to load balancing.
\end{proposition}

\lbsCommunication*
\begin{proof}
Apply \cref{prop:lbs-communication} to the sparsifiers of \cref{thm:improved-sparsifier},
with \(m=O(n\log^3 n)\) and \(\alpha=O(n^{1/3-\varepsilon_0})\).
\end{proof}

\subsection{Proof of the Normalization Lemma}
\label{app:clean}

\begin{proof}[Proof of \cref{lem:clean}]
    Let \(m_0:=|E(F)|\) and \(\ell:=\lceil\log_2 N\rceil+1\).  Bucket the
    matchings by dyadic size, and choose a bucket \([t,2t)\) containing at
    least \(m_0/\ell\) edges.  Within this bucket, retain whole matchings
    until their total number of edges first reaches \(m_0/\ell\).  Since one
    matching has at most \(N\) edges, the retained edge count \(m\) satisfies
    \[
        \frac{m_0}{\ell}
        \le m
        \le \frac{m_0}{\ell}+N
        =
        \Theta\left(\frac{m_0}{\ell}\right).
    \]
    The final equality uses \(\mathsf d_F\ge C\log N\).  Removing whole
    matchings preserves the \(\beta\)-matching-contractor property, and
    every remaining matching has size in \([t,2t)\).

    Fix a sufficiently small absolute constant \(\eta>0\), and set
    \(D:=\eta m/N=\Theta(\mathsf d_F/\log N)\).  Taking \(C\) large enough,
    we may assume \(D\ge2\).

    We first prepare the left degrees.  Delete every left vertex of degree
    less than \(D\).  This removes at most \(ND=\eta m\) edges.  For each
    remaining left vertex, partition its incident edges
    into groups of sizes between \(D\) and \(2D\), and replace the vertex
    by one copy for each group.  Because an original matching contains at
    most one edge at that vertex, its surviving edges still form a
    matching after the split.  There are at most \(m/D=N/\eta\) left copies.

    Repeatedly apply any applicable rule below; the scales \(D\) and \(t\)
    remain fixed.
    \begin{enumerate}
        \item \emph{Low left degree.}
        If a left vertex has current degree less than \(\eta D\), delete
        that vertex and its incident edges.
        \item \emph{Small matching.}
        If a current matching \(M_i\) has fewer than \(\eta t\) edges,
        delete its remaining edges and discard it.
        \item \emph{Sparse residual neighborhood.}
        With \(L_i,R_i,H_i\) defined in the current graph, if
        \(r\in R_i\) and
        \(\deg_{H_i}(r)<\eta\beta\), delete every edge of \(H_i\)
        incident to \(r\).
    \end{enumerate}
    We now bound the number of deleted edges.  The initial deletion costs at most
    \(\eta m\).  During the repeated process, each of the at most \(m/D\)
    left copies is deleted once and has fewer than \(\eta D\) edges, so the
    first rule costs at most another \(\eta m\).

    If \(q\) matchings were retained from the chosen bucket, then
    \(qt\le m\).  Each matching discarded by the second rule has fewer
    than \(\eta t\) remaining edges, so that rule costs at most
    \(q\eta t\le\eta m\).

    Let \(M_i^0\) and \(R_i^0\) denote a matching and its residual
    neighborhood just after the dyadic bucket was chosen.  Splitting left
    vertices and deleting edges cannot create a residual neighbor, so
    \[
        \sum_i|R_i^0|
        \le
        \sum_i\frac{|M_i^0|}{\beta}
        =
        \frac{m}{\beta}.
    \]
    A pair \((i,r)\) is processed by the third rule at most once: afterward
    \(r\notin R_i\), and later deletions cannot make it reappear.  The rule
    removes fewer than \(\eta\beta\) edges for each such pair, so its total
    cost is at most \(\eta\beta\sum_i|R_i^0|\le\eta m\).
    Altogether, at most \(4\eta m\) edges are removed.  Taking
    \(\eta<1/8\) leaves \(\Theta(m)\) edges.

    At termination, every surviving left vertex has degree between
    \(\eta D\) and \(2D\), and every surviving matching has size between
    \(\eta t\) and \(2t\).  Thus the left degrees are comparable and all
    matching sizes are within an absolute factor of their average.  The
    third rule gives \(\deg_{H_i}(r)\ge\eta\beta\) for every \(r\in R_i\).

    It remains to check the contraction parameter.  For every surviving
    matching, its final residual neighborhood is a subset of \(R_i^0\).
    Therefore
    \[
        |R_i|
        \le |R_i^0|
        \le \frac{|M_i^0|}{\beta}
        <
        \frac{2t}{\beta}
        \le
        \frac{2}{\eta\beta}|M_i|.
    \]
    Hence the final graph is an \(\alpha\)-matching-contractor for
    \(\alpha:=\eta\beta/2=\Omega(\beta)\).

    Remove isolated left vertices, and let \(n\) be the number of remaining
    left vertices.  The final graph has \(\Theta(m)\) edges and every left
    degree is at most \(2D\), so \(n=\Omega(m/D)=\Omega(N)\).
    There are at most \(m/D=N/\eta\) left copies, so \(n=\Theta(N)\), and
    therefore
    \[
        \mathsf d_G
        =
        \Theta(m/N)
        =
        \Theta\left(\frac{\mathsf d_F}{\log N}\right).
    \]
    Since \(D=\Theta(\mathsf d_G)\), these bounds give all the conclusions of
    the lemma.
\end{proof}

\section{Streaming Algorithm for Load-Balancing Sparsifiers}
\label{sec:streaming-sparsifiers}

In this section, we prove the streaming upper bound stated in the introduction.

\lbsStreamingSparsifier*

As explained in the technical overview, our algorithm replaces each semi-matching in the
two-level construction of~\cite{konrad2016approximating} with a subgraph that stores $O(\log n)$ edges per left vertex. We first
describe the rule for storing these edges and the buckets it defines. We then apply the same
rule inside each bucket. The main point of the analysis is that an edge we fail to store has
\emph{two blockers}, one at each level. These blockers give the two charging inequalities
needed for the $\widetilde O(n^{1/3})$ guarantee.

\subsection{The two-level bucketing algorithm}
\label{subsec:streaming-algorithm}

\paragraph{The first level: storing edges according to their right endpoint's degree.}
Let $G=(L\cup R,E)$ be the input graph, with $n=|L|$. At the first level, we maintain a
subgraph $H_1\subseteq G$, initially empty, and track the degree of each right vertex in this
stored subgraph:
\begin{equation}
  d_1(r):=\deg_{H_1}(r),\qquad r\in R.
  \label{eq:first-counter-invariant}
\end{equation}
Thus $d_1(r)$ increases by one whenever we store a new edge incident to $r$ at the first level.

To limit how many edges we store at each left vertex, we use a table $P_1$ with one row per
$a\in L$ and $\lambda$ columns, where
\[
  \lambda:=\ceil{\log_2(n+1)}
  \qquad\text{and}\qquad
  \mathcal I:=\{0,1,\ldots,\lambda-1\}.
\]
Each entry $P_1[a,i]$ is initially empty, denoted by $\bot$, and can hold one edge incident
to $a$. The edges in the table are precisely the edges of $H_1$.

When an edge $(a,s)$ arrives, its first-level update uses the current degree of $s$ to choose
the column
\[
  i:=\floor{\log_2(d_1(s)+1)},
  \qquad\text{so that}\qquad
  2^i\le d_1(s)+1<2^{i+1}.
\]
In other words, each column corresponds to a range of right degrees: column $0$ is used for
degree $0$, column $1$ for degrees $1$ and $2$, column $2$ for degrees $3$ through $6$, and
so on. If $P_1[a,i]$ is empty, we store $(a,s)$ there and increase $d_1(s)$ by one. Otherwise,
we keep the edge already in that entry and say that it \emph{blocks} $(a,s)$ at the first
level. For example, an earlier edge $(a,r)$ can block $(a,s)$ because both request the same
column in $a$'s row. An edge that has been stored is never removed or moved to another column.

\paragraph{The buckets created by the first level.}
For each $r\in R$ and $i\in\mathcal I$, define
\[
  B_{r,i}:=\{a\in L:P_1[a,i]=(a,r)\}.
\]
This bucket consists of the left vertices whose edges to $r$ were stored in column $i$.
Storing $(a,r)$ in that column makes $a$ a permanent member of $B_{r,i}$. A left vertex can
belong to several buckets, but to at most one for each column, and hence to at most $\lambda$
buckets in total.

These buckets connect the storage rule to the analysis. All edges defining $B_{r,i}$ were
stored while $d_1(r)+1$ was between $2^i$ and $2^{i+1}-1$, and each such insertion increased
$d_1(r)$ by one. Consequently, $|B_{r,i}|\le 2^i$. If an edge $(a,s)$ is blocked in column
$i$ by $(a,r)$, the same index was chosen using the degree of $s$, giving
\[
  |B_{r,i}|\le 2^i\le d_1(s)+1.
\]
\emph{This is the key comparison: the size of the blocker's bucket is bounded by the degree
of the rejected edge's right endpoint.} It will allow us to sum bucket sizes over a matching.

\paragraph{The second level: applying the same rule inside each bucket.}
For each first-level bucket $B_{r,i}$, we maintain a separate table $P_2^{r,i}$ with one row
per current member of $B_{r,i}$ and the same $\lambda$ columns. Let $H_2^{r,i}$ consist of
the edges stored in this table, and track
\begin{equation}
  d_2^{r,i}(s):=\deg_{H_2^{r,i}}(s),\qquad s\in R.
  \label{eq:second-counter-invariant}
\end{equation}
Each second-level instance therefore has its own degree counters, measured in its own stored
subgraph.

When $a$ joins $B_{r,i}$, we give it an empty row in $P_2^{r,i}$. Every \emph{later} edge
$(a,s)$ is processed by this table: compute
\[
  j:=\floor{\log_2(d_2^{r,i}(s)+1)},
\]
store $(a,s)$ in $P_2^{r,i}[a,j]$ if that entry is empty, and increase $d_2^{r,i}(s)$ by one
when the edge is stored. If the entry already contains an edge, that edge blocks $(a,s)$ in
this second-level instance. The right endpoint $s$ can be any vertex of $R$; the restriction
to $B_{r,i}$ concerns only the left endpoint. Edges that arrived before $a$ joined the bucket
are not processed again.

\paragraph{Putting the two levels together.}
On the arrival of $(a,s)$, we first process it in the second-level table of every bucket
that already contains $a$. We then perform its first-level update, creating a new
second-level row if this update places $a$ in a new bucket. Thus the edge that creates a
membership is stored at the first level, and the associated second-level row begins processing
edges after that membership is created.

The important consequence of this order is that a first-level blocker identifies a
second-level table that has already processed the arriving edge. Indeed, if $P_1[a,i]$
already contains $(a,r)$, then $a$ already belongs to $B_{r,i}$. If $(a,s)$ is omitted from
the output, it must also have been blocked in $P_2^{r,i}$.

The output is the subgraph
\[
  H:=H_1\cup\bigcup_{r\in R,\,i\in\mathcal I}H_2^{r,i}.
\]
An edge stored in several tables contributes a single edge to this union. All degree counters
default to zero. The complete algorithm appears in \cref{fig:streaming-two-level-bucketing}.

\begin{figure}[t]
  \centering
  \setlength{\fboxsep}{8pt}
  \fbox{\begin{minipage}{0.92\linewidth}
    \textsc{Two-Level-Bucketing}

    \smallskip
    \textbf{Initialization.}
    Set $P_1[a,i]\gets\bot$ for all $a\in L$ and $i\in\mathcal I$.
    All degree counters default to zero. A second-level table is created when its bucket
    receives its first member.

    \smallskip
    \textbf{On the arrival of an edge $(a,s)$:}
    \begin{tabbing}
      \qquad\=\qquad\=\qquad\=\kill
      1. \textit{Process the edge inside the buckets that already contain $a$.}\\
      \> \textbf{for each} $i_0\in\mathcal I$ with $P_1[a,i_0]\ne\bot$ \textbf{do}\\
      \>\> let $(a,r)$ be the edge in $P_1[a,i_0]$\\
      \>\> $j\gets\floor{\log_2(d_2^{r,i_0}(s)+1)}$\\
      \>\> \textbf{if} $P_2^{r,i_0}[a,j]=\bot$ \textbf{then}\\
      \>\>\> $P_2^{r,i_0}[a,j]\gets(a,s)$\\
      \>\>\> $d_2^{r,i_0}(s)\gets d_2^{r,i_0}(s)+1$\\[2pt]
      2. \textit{Process the edge at the first level.}\\
      \> $i\gets\floor{\log_2(d_1(s)+1)}$\\
      \> \textbf{if} $P_1[a,i]=\bot$ \textbf{then}\\
      \>\> $P_1[a,i]\gets(a,s)$\\
      \>\> $d_1(s)\gets d_1(s)+1$\\
      \>\> initialize an empty row $P_2^{s,i}[a,\cdot]$ of $\lambda$ entries
    \end{tabbing}

    \textbf{Output.}
    The graph $H$ formed by the edges stored in all tables.
  \end{minipage}}
  \caption{The two-level bucketing algorithm. Each table stores edges and maintains the
  right degrees in its own stored subgraph.}
  \label{fig:streaming-two-level-bucketing}
\end{figure}

\subsection{Analysis}
\label{subsec:streaming-analysis}

After bounding the space, we show that every set $X\subseteq L$ that is matchable in $G$
retains at least $|X|/\widetilde O(n^{1/3})$ neighbors in $H$. Matchable sets are useful here
because a load-$q$ assignment can be split into $q$ matchings. We use this observation and
Hall's theorem at the end to obtain the load-balancing guarantee.

Throughout the analysis, stored subgraphs and buckets refer to their contents at the end of
the stream. Whenever we use a degree at an edge's arrival, we say so explicitly.

\paragraph{Space and bucket sizes.}

\begin{lemma}[Space and output size]
\label{lem:streaming-size}
The algorithm uses $O(n\lambda^2)$ words of $O(\log n)$ bits and outputs at most
$n\lambda+n\lambda^2$ edges.
\end{lemma}

\begin{proof}
The table $P_1$ has $n\lambda$ entries. Every filled entry creates one second-level row of
$\lambda$ entries, so all second-level tables together have at most $n\lambda^2$ entries.
These tables represent all stored subgraphs and bucket memberships.

We store a degree counter only when it becomes positive. The number of such counters is at
most the number of occupied table entries, since each positive counter counts at least one
edge in its table. Thus the counters also use $O(n\lambda^2)$ words. Vertex identifiers use
$O(\log n)$ bits by our streaming model, and the indices, degrees, and table pointers fit in
the same word size. Finally, every output edge occurs in a table entry, giving the stated
bound on $|E(H)|$.
\end{proof}

We will use the bucket-size observation from the algorithm description at both levels.

\begin{lemma}[Bucket sizes]
\label{lem:streaming-bucket-sizes}
For every $r\in R$ and $i\in\mathcal I$, we have $|B_{r,i}|\le 2^i$. Similarly, in any
fixed second-level table, at most $2^j$ rows contain an edge to a given right vertex $s$ in
column $j$.
\end{lemma}

\begin{proof}
In either table, an edge to $s$ is stored in column $j$ only when the current degree of $s$
in that table's subgraph lies in
\[
  \{2^j-1,2^j,\ldots,2^{j+1}-2\}.
\]
There are $2^j$ such degree values. Each insertion increases the degree by one, so at most
$2^j$ insertions can use this column for $s$.
\end{proof}

\paragraph{Following the matching edges that were not stored.}
Fix a set $X\subseteq L$ that is matchable in $G$, and a matching that covers $X$. Write
$(a,u_a)$ for the matching edge of $a\in X$. Let
\[
  T:=N_H(X),\qquad k:=|T|,
  \qquad\text{and}\qquad
  X_{\mathrm{out}}:=\{a\in X:u_a\notin T\}.
\]
The matching partners $u_a$ are distinct, so at most $k$ of them lie in $T$. Therefore
\begin{equation}
  |X|\le k+|X_{\mathrm{out}}|.
  \label{eq:Xsplit}
\end{equation}
Our task is to bound $|X_{\mathrm{out}}|$: these vertices had distinct available partners in
$G$, yet all their stored neighbors lie in $T$.

For $a\in X_{\mathrm{out}}$, consider the arrival of $(a,u_a)$. Let $i_a$ be its chosen
first-level column. Since $u_a\notin T$, the algorithm did not store this edge. Hence
$P_1[a,i_a]$ already contained an edge $(a,r_a)$, which is its first-level blocker. We have
$r_a\in T$ and $a\in B_{r_a,i_a}$.

Assign each $a\in X_{\mathrm{out}}$ to this particular bucket by defining
\[
  X_{r,i}:=\{a\in X_{\mathrm{out}}:(r_a,i_a)=(r,i)\},
  \qquad x_{r,i}:=|X_{r,i}|.
\]
Although the buckets themselves may overlap, the sets $X_{r,i}$ partition
$X_{\mathrm{out}}$. Each satisfies $X_{r,i}\subseteq B_{r,i}$, and only the $k\lambda$
indices in $T\times\mathcal I$ can occur.

The two levels now give complementary bounds. The first level shows that few vertices
of $X_{\mathrm{out}}$ can be assigned to large buckets. The second level shows that a bucket
assigned many of these vertices must be large. Combining the bounds will limit the total
number of vertices in $X_{\mathrm{out}}$.

\paragraph{First-level charging: few vertices can select large buckets.}

\begin{lemma}[First-level charging inequality]
\label{lem:streaming-first-charge}
\begin{equation}
  \sum_{r\in T}\sum_{i\in\mathcal I}x_{r,i}|B_{r,i}|
  \le n(\lambda+1).
  \label{eq:firstcharge}
\end{equation}
\end{lemma}

\begin{proof}
We charge each $a\in X_{\mathrm{out}}$ the size of the bucket containing its first-level
blocker. If $a\in X_{r,i}$, then $|B_{r,i}|\le 2^i$ by
\cref{lem:streaming-bucket-sizes}. On the arrival of $(a,u_a)$, column $i$ was chosen using
$d_1(u_a)$, so $2^i\le d_1(u_a)+1$ at that time. Degrees only increase, and thus
\[
  |B_{r,i}|\le 2^i\le \deg_{H_1}(u_a)+1.
\]
Summing over $X_{\mathrm{out}}$, we obtain
\[
  \sum_{r\in T}\sum_{i\in\mathcal I}x_{r,i}|B_{r,i}|
  \le \sum_{a\in X_{\mathrm{out}}}\bigl(\deg_{H_1}(u_a)+1\bigr)
  \le |E(H_1)|+|X_{\mathrm{out}}|
  \le n(\lambda+1).
\]
The middle inequality uses the distinctness of the matching partners: each right degree is
counted at most once.
\end{proof}

This is the charging argument from the technical overview, with one bucket chosen for each
rejected matching edge. We next apply the same comparison inside a fixed bucket.

\paragraph{Second-level charging: many vertices force a large bucket.}

\begin{lemma}[Second-level bound]
\label{lem:streaming-second-charge}
\label{lem:streaming-local-square}
For every $(r,i)\in T\times\mathcal I$,
\begin{equation}
  x_{r,i}^2\le k\lambda(\lambda+1)|B_{r,i}|.
  \label{eq:square}
\end{equation}
\end{lemma}

\begin{proof}
Fix one bucket $B:=B_{r,i}$, and abbreviate its second-level table and stored subgraph by
$P:=P_2^{r,i}$ and $F:=H_2^{r,i}$. We will count the vertices of $X_{r,i}$ according to
their blockers in this table.

For each $a\in X_{r,i}$, the edge $(a,r)$ was already stored when $(a,u_a)$ arrived.
Consequently, $a$ was already a member of $B$, and $P$ processed $(a,u_a)$. This edge was
blocked here as well, since otherwise it would belong to $H$. Let $j_a$ be the column it
requested and $(a,s_a)$ the edge already stored there. This second-level blocker belongs to
$H$, so $s_a\in T$.

For $s\in T$ and $j\in\mathcal I$, let
\[
  y_{s,j}:=\bigl|\{a\in X_{r,i}:(s_a,j_a)=(s,j)\}\bigr|.
\]
These counts partition the vertices under consideration:
\begin{equation}
  x_{r,i}=\sum_{s\in T}\sum_{j\in\mathcal I}y_{s,j}.
  \label{eq:second-partition}
\end{equation}
All $y_{s,j}$ vertices have an edge to $s$ stored in column $j$ of $P$, so
$y_{s,j}\le 2^j$ by \cref{lem:streaming-bucket-sizes}. For each such vertex $a$, column
$j$ was chosen from the degree of $u_a$ in this same second-level instance. Comparing with
the final degree gives
\[
  y_{s,j}\le 2^j\le \deg_F(u_a)+1.
\]
We charge every vertex counted by $y_{s,j}$ an amount $y_{s,j}$. Summing these
charges produces $y_{s,j}^2$, so the last inequality gives
\begin{equation}
  \sum_{s\in T}\sum_{j\in\mathcal I}y_{s,j}^2
  \le \sum_{a\in X_{r,i}}\bigl(\deg_F(u_a)+1\bigr)
  \le |E(F)|+x_{r,i}
  \le |B|(\lambda+1).
  \label{eq:secondcharge}
\end{equation}
Here the matching partners are distinct, $P$ has $|B|\lambda$ entries, and
$x_{r,i}\le |B|$.

Finally, there are only $k\lambda$ possible pairs $(s,j)$. Applying Cauchy--Schwarz to
\cref{eq:second-partition} and then using \cref{eq:secondcharge}, we get
\[
  x_{r,i}^2
  \le k\lambda\sum_{s\in T}\sum_{j\in\mathcal I}y_{s,j}^2
  \le k\lambda(\lambda+1)|B|.
\]
\end{proof}

\paragraph{Combining the two bounds.}
The second-level bound gives a lower bound on $|B_{r,i}|$ in terms of $x_{r,i}^2$.
The first-level inequality multiplies this bucket size by $x_{r,i}$.

\begin{lemma}[Expansion guarantee]
\label{lem:streaming-expansion}
\label{lem:streaming-escaping}
For every $X\subseteq L$ that is matchable in $G$,
\[
  |N_H(X)|\ge \frac{|X|}{\ceil{4n^{1/3}\lambda^{5/3}}}.
\]
\end{lemma}

\begin{proof}
Use the notation above. Multiplying \cref{eq:square} by $x_{r,i}$, summing over the
first-level indices, and applying \cref{eq:firstcharge} gives
\[
  \sum_{r\in T}\sum_{i\in\mathcal I}x_{r,i}^3
  \le k\lambda(\lambda+1)
      \sum_{r\in T}\sum_{i\in\mathcal I}x_{r,i}|B_{r,i}|
  \le nk\lambda(\lambda+1)^2.
\]
There are at most $k\lambda$ counts $x_{r,i}$. H\"older's inequality therefore gives
\[
  |X_{\mathrm{out}}|^3
  =\left(\sum_{r\in T}\sum_{i\in\mathcal I}x_{r,i}\right)^3
  \le (k\lambda)^2\sum_{r\in T}\sum_{i\in\mathcal I}x_{r,i}^3
  \le nk^3\lambda^3(\lambda+1)^2
  \le 4nk^3\lambda^5.
\]
Thus
\begin{equation}
  |X_{\mathrm{out}}|\le 4^{1/3}n^{1/3}k\lambda^{5/3}.
  \label{eq:escape}
\end{equation}
Together with \cref{eq:Xsplit}, this yields
\[
  |X|\le k+4^{1/3}n^{1/3}k\lambda^{5/3}
  \le 4n^{1/3}\lambda^{5/3}k,
\]
which proves the claimed expansion bound.
\end{proof}

\paragraph{From expansion to load balancing.}
We finish by showing that the expansion guarantee preserves load balancing for every subset
of left vertices.

\begin{proof}[Proof of \cref{thm:streaming-sparsifier-restatement}]
The algorithm reads each stream edge once. By \cref{lem:streaming-size}, it uses
$O(n\lambda^2\log n)$ bits and stores $O(n\lambda^2)$ edges. Set
\[
  A:=\ceil{4n^{1/3}\lambda^{5/3}}.
\]
We show that $H$ is an $A$-load-balancing sparsifier.

Fix a nonempty $Y\subseteq L$ and let $q:=\operatorname{\textsc{opt}}(G[Y\cup R])$. Split an optimal load-$q$
assignment into $q$ matchings, and let $Y_1,\ldots,Y_q$ be the resulting partition of $Y$
into sets that are matchable in $G$. For any $Z\subseteq Y$, one of these sets satisfies
$|Z\cap Y_t|\ge |Z|/q$. Applying \cref{lem:streaming-expansion} to $Z\cap Y_t$ gives
\[
  |N_H(Z)|\ge |N_H(Z\cap Y_t)|
  \ge \frac{|Z\cap Y_t|}{A}
  \ge \frac{|Z|}{Aq}.
\]
By the capacitated form of Hall's theorem, an assignment of $Y$ with maximum load $Aq$
exists whenever $|Z|\le Aq|N_H(Z)|$ for all $Z\subseteq Y$. Hence
\[
  \operatorname{\textsc{opt}}(H[Y\cup R])\le Aq
  =A\operatorname{\textsc{opt}}(G[Y\cup R]).
\]
Since $Y$ was arbitrary, this proves the sparsifier guarantee. Substituting
$\lambda=O(\log n)$ gives the theorem's stated space, output-size, and approximation bounds.
\end{proof}

The streaming algorithm for load balancing follows by solving the problem on the stored
sparsifier. We restate the corollary for convenience.

\lbsStreamingLoadBalancing*

\begin{proof}
Compute the sparsifier $H$ from \cref{thm:streaming-sparsifier-restatement}, then compute an
optimal load-balancing assignment in $H$. The sparsifier guarantee makes this a
$\widetilde O(n^{1/3})$-approximation for $G$. After omitting isolated right vertices,
$H$ has at most $n+|E(H)|$ vertices; a capacitated max-flow computation finds the assignment
using $O(n+|E(H)|)$ words. Thus the postprocessing also fits in $\widetilde O(n)$ space.
\end{proof}

\section{Matching Contractor Lower Bound}
\label{sec:matching-contractor-lower-bound}

Our goal in this section is to prove \Cref{thm:linear-size-lower}, which we
restate below.

\begin{boxedtheorem}
\lbsLinearSizeLower*
\end{boxedtheorem}

By the duality between load-balancing sparsifiers and matching contractors,
it is enough to construct a matching contractor with $n$ vertices on the
left, contraction parameter $n^{1/4+1/c}$, and $\Omega(cn)$ edges.  See
\Cref{subsec:LBS-over} for an overview of matching contractors and their
equivalence with load-balancing sparsifiers.

We now construct the matching contractors that give this lower bound.  Fix
an integer $t\ge2$.  The construction has $t$ rounds, following the framework
of \cite{assadi2025streaming}, and each round contributes one edge incident
to every left vertex.  Set
\[
    h:=\floor{t/2}
    \qquad\text{and}\qquad
    K:=\binom{t}{h}.
\]
For every $q=3^a$ with $a\ge1$, define a bipartite graph
$H_q=(L_q\cup R_q,E_q)$ with
\[
    L_q:=\mathbb{F}_q^K\times\mathbb{F}_q^K
    \qquad\text{and}\qquad
    R_q:=[t]\times\mathbb{F}_q^K.
\]
The left side has $q^{2K}$ vertices.  The right side has $tq^K$ vertices and
is divided into $t$ components: a right vertex $(i,z)$ belongs to component $i$.
The parameter $q$ controls the size of the graph, whereas $t$ remains fixed.

We next specify the one neighbor that each left vertex receives in each
right-side component.  Let
\[
    \mathcal{B}:=\{B\subseteq[t]:|B|=h\}.
\]
Note that $|\mathcal{B}|=K$. We use the sets $B\in\mathcal{B}$ to label the $K$
coordinates of every vector in $\mathbb{F}_q^K$.  For each round $i\in[t]$,
define $\sigma_i\in\{\pm1\}^{\mathcal{B}}$ by
\[
    \sigma_i(B):=
    \begin{cases}
        +1,&i\in B,\\
        -1,&i\notin B.
    \end{cases}
\]
Thus, for a fixed coordinate $B\in\mathcal{B}$, the $B$-coordinate of
$\sigma_i$ is positive exactly in the rounds $i\in B$.  In particular, each
coordinate is positive in $h$ of the $t$ sign vectors and negative in the
other $t-h$ sign vectors.  For example, if $t=4$ and $B=\{1,3\}$, then the
$B$-coordinate is positive in $\sigma_1,\sigma_3$ and negative in
$\sigma_2,\sigma_4$.

All vector arithmetic below is over $\mathbb{F}_q$.  For a left vertex
$(x,y)\in L_q$, let $x_B$ and $y_B$ denote the coordinates indexed by
$B\in\mathcal{B}$, and define $F_i(x,y)\in\mathbb{F}_q^K$ by
\[
    \bigl(F_i(x,y)\bigr)_B:=y_B+\sigma_i(B)x_B
    \qquad\text{for every $B\in\mathcal{B}$}.
\]
In round $i$, add
$\bigl((x,y),(i,F_i(x,y))\bigr)$ to $H_q$ for every $(x,y)\in L_q$.
Consequently, every left vertex has one incident edge in each round.  The
round index in the right endpoint keeps the right-side components disjoint.

\begin{proposition}[Size]
\label{prop:lower-size}
The graph $H_q$ is simple and has
\[
    |L_q|=q^{2K}=:n_q,
    \qquad
    |R_q|=tq^K=t\sqrt{n_q},
    \qquad
    |E_q|=tn_q.
\]
\end{proposition}

\begin{proof}
We add one edge to $H_q$ for each vertex in $L_q$ in every round $i \in [t]$,  so
$|E_q|=t|L_q|=tn_q$.  Two edges created from different left vertices have
different left endpoints, while two edges created in different rounds have
right endpoints in different components.  Thus no edge is repeated.  The vertex
counts follow from the definitions of $L_q$ and $R_q$.
\end{proof}

The edges we add to $H_q$ within one round do not  form a matching, so we divide
them into matchings using a second linear label.  For $i\in[t]$, define
$G_i:L_q\to\mathbb{F}_q^K$ by
\[
    \bigl(G_i(x,y)\bigr)_B:=y_B-\sigma_i(B)x_B
    \qquad\text{for every $B\in\mathcal{B}$}.
\]
For $z\in\mathbb{F}_q^K$, let
\[
    L_{i,z}:=\{(x,y)\in L_q:G_i(x,y)=z\}
\]
and define the corresponding set of round-$i$ edges by
\[
    M_{i,z}:=
    \bigl\{
        \bigl((x,y),(i,F_i(x,y))\bigr):(x,y)\in L_{i,z}
    \bigr\}.
\]
We now verify that the edge sets $M_{i,z}$ are indeed matchings and that they
partition $E_q$.

\begin{lemma}[Matching partition]
\label{lem:lower-matching-partition}
The sets
\[
    \{M_{i,z}:i\in[t],\ z\in\mathbb{F}_q^K\}
\]
partition $E_q$ into matchings of size $q^K$.
\end{lemma}

\begin{proof}
Every edge added in round $i$ has a unique label $z=G_i(x,y)$, so the sets
$M_{i,z}$ partition $E_q$. 
It remains to check that each set is a matching.  Fix $i$, and
consider the map
\[
    (F_i,G_i):L_q\longrightarrow
    \mathbb{F}_q^K\times\mathbb{F}_q^K.
\]
Suppose $(F_i(x,y),G_i(x,y))=(f,g)$.  Because $q$ is odd, $2$ is invertible
in $\mathbb{F}_q$, and each coordinate of $(x,y)$ can be recovered from
\[
    x_B=\tfrac12\sigma_i(B)(f_B-g_B),
    \qquad
    y_B=\tfrac12(f_B+g_B).
\]
Hence $(F_i,G_i)$ is a bijection.  After fixing $G_i(x,y)=z$, the value
$F_i(x,y)$ therefore ranges exactly once over $\mathbb{F}_q^K$.  The edges
of $M_{i,z}$ have distinct left endpoints and distinct right endpoints, so
$M_{i,z}$ is a matching of size $q^K$.
\end{proof}

We still need to bound the residual neighborhood of each $M_{i,z}$.  The
key quantity is the number of coordinates on which the sign vectors for two
rounds disagree.  The Hamming distance between $\sigma_i$ and $\sigma_j$ is
\[
    d_{\mathrm H}(\sigma_i,\sigma_j)
    :=|\{B\in\mathcal{B}:\sigma_i(B)\ne\sigma_j(B)\}|.
\]
When we examine the left endpoints of a round-$i$ matching in another round
$j$, every coordinate on which $\sigma_i$ and $\sigma_j$ disagree will
cancel.  Thus a larger Hamming distance will give a smaller residual
neighborhood.  The following lemma shows that every two distinct sign
vectors $\sigma_i$ and $\sigma_j$ have large Hamming distance.  This bound is
tight by the Plotkin bound in coding theory.

\begin{lemma}[Hamming distance of the sign vectors]
\label{lem:lower-sign-distance}
Every two distinct sign vectors $\sigma_i$ and $\sigma_j$ have the same
Hamming distance
\[
    d_{\mathrm H}(\sigma_i,\sigma_j)
      =D:=2\binom{t-2}{h-1}
      =\frac{K\floor{t^2/4}}{\binom{t}{2}}.
\]
\end{lemma}

\begin{proof}
Fix distinct $i,j\in[t]$.  The sign vectors $\sigma_i$ and $\sigma_j$ differ at coordinate
$B\in\mathcal{B}$ exactly when $B$ contains one of $i,j$ but not the other.
There are $2\binom{t-2}{h-1}$ such coordinates, independently of the pair.
Finally,
\[
    2\binom{t-2}{h-1}
    =
    K\cdot\frac{2h(t-h)}{t(t-1)}
    =
    \frac{K\floor{t^2/4}}{\binom{t}{2}},
\]
where the last equality uses $h(t-h)=\floor{t^2/4}$.
\end{proof}

We next translate the Hamming-distance calculation into the number of
neighbors contributed by one other round.

\begin{lemma}[Neighbors in another round]
\label{lem:lower-cross-round}
Fix distinct rounds $i,j\in[t]$ and a label $z\in\mathbb{F}_q^K$.  The set
of round-$j$ neighbors of $L_{i,z}$ is
\[
    \{j\}\times F_j(L_{i,z}),
\]
and it has size $q^{K-D}$.
\end{lemma}

\begin{proof}
For $(x,y)\in L_{i,z}$, the equation $G_i(x,y)=z$ determines
$y_B=z_B+\sigma_i(B)x_B$ in every coordinate $B\in\mathcal{B}$.  Therefore,
\[
    \bigl(F_j(x,y)\bigr)_B
    =
    z_B+\bigl(\sigma_i(B)+\sigma_j(B)\bigr)x_B.
\]
If $\sigma_i(B)\ne\sigma_j(B)$, then the coefficient of $x_B$ is zero, so
the $B$-coordinate of every image is fixed to $z_B$.  If
$\sigma_i(B)=\sigma_j(B)$, then the coefficient is $+2$ or $-2$ and is
nonzero in $\mathbb{F}_q$.  In this case the image coordinate can take every
value in $\mathbb{F}_q$.  Indeed, after $z$ is fixed, every choice of $x$
determines a unique $y$, so the coordinates of $x$ vary independently.

By \Cref{lem:lower-sign-distance}, the signs disagree in $D$ coordinates and
agree in $K-D$ coordinates.  Hence $D$ image coordinates are fixed and the
other $K-D$ coordinates are free, giving exactly $q^{K-D}$ round-$j$
neighbors.
\end{proof}

We are now ready to prove that $H_q$ is a matching contractor.

\begin{proposition}[Contraction property]
\label{prop:lower-contraction}
For every $i\in[t]$ and $z\in\mathbb{F}_q^K$,
\[
    \left|
        N_{H_q\setminus M_{i,z}}\bigl(L(M_{i,z})\bigr)
    \right|
    =(t-1)q^{K-D}.
\]
Consequently, $H_q$ is a matching contractor with parameter
$q^D/(t-1)$.
\end{proposition}

\begin{proof}
By \Cref{lem:lower-matching-partition}, the sets $M_{i,z}$ form a matching
partition of $E_q$.  Fix $M_{i,z}$.  By definition, its left endpoints are
the vertices of $L_{i,z}$.  Each such vertex has only one edge in round $i$,
and that edge belongs to $M_{i,z}$.  Thus removing $M_{i,z}$ removes every
round-$i$ edge incident to one of its left endpoints.

For each $j\ne i$, \Cref{lem:lower-cross-round} identifies the round-$j$
neighbor set as $\{j\}\times F_j(L_{i,z})$ and gives it size $q^{K-D}$.
Neighbor sets from different rounds are disjoint because their right
vertices have different first coordinates.  Summing over the $t-1$
remaining rounds gives the displayed residual-neighborhood size.  Since
$|M_{i,z}|=q^K$,
\[
    \frac{|M_{i,z}|}
         {|N_{H_q\setminus M_{i,z}}(L(M_{i,z}))|}
    =
    \frac{q^K}{(t-1)q^{K-D}}
    =
    \frac{q^D}{t-1}.
\]
This is precisely the matching contractor condition with parameter
$q^D/(t-1)$.
\end{proof}

It remains to express the contraction parameter of $H_q$ in terms of the number of left
vertices.  Because $n_q=q^{2K}$, we have
$q^D=n_q^{D/(2K)}$.  By \Cref{lem:lower-sign-distance},
\[
    \beta_t:=\frac{D}{2K}
    =\frac{h(t-h)}{t(t-1)}
    \ge\frac14+\frac{1}{4t}.
\]
Thus \Cref{prop:lower-contraction} gives $H_q$ contraction parameter
$n_q^{\beta_t}/(t-1)$. Then for every fixed $t$ and all sufficiently
large $q$,
\[
    \frac{n_q^{\beta_t}}{t-1}
    \ge \Omega(n_q^{1/4+1/(4t)}).
\]

\begin{proof}[Proof of \Cref{thm:linear-size-lower}]
By
\Cref{prop:lower-size,lem:lower-sign-distance,prop:lower-contraction}, for
every fixed integer $t\ge2$ and infinitely many $n$, there is a matching
contractor with $n$ vertices on the left, at least $tn/2$ edges, and
contraction parameter at least $n^{1/4+1/(4t)} / t$. Applying the duality between load-balancing sparsifiers and matching
contractors in \Cref{lem:MC_duality} 
and re-parameterizing 
implies that for every $c > 1$, there exists infinitely many $n$-client assignment graphs such that $n^{1/4+1/c}$-load-balancing sparsifiers
require $\Omega(cn)$ edges. 
This completes the proof of
\Cref{thm:linear-size-lower}.
\end{proof}

\section{AI Methods}

\label{sec:ai_methods}

In this section, we provide a more detailed description of how we used AI tools in the development of each of our results.

\paragraph{\Cref{thm:improved-sparsifier}.} This theorem was proved without the use of AI tools. We attempted to improve the approximation to $O(n^{1/4})$ by running  ChatGPT 5.6 Sol on Ultra mode in Codex for 48 hours, but this was unsuccessful. 
ChatGPT 5.6 Sol suggested an  argument to improve our bounds by recursively applying them, but the improvement was extremely small.

\paragraph{\Cref{thm:streaming-sparsifier}.} This theorem was proved in essentially one shot by ChatGPT 5.5 Pro using the web version of ChatGPT. In our prompt, we included some of our intuition about the problem; it is unclear whether ChatGPT used this intuition in its solution.

\paragraph{\Cref{thm:linear-size-lower}.}
This theorem was proved in one shot by ChatGPT 5.6 Sol on Ultra mode. We initially asked ChatGPT to prove that there exists $O(n)$-edge $O(n^{1/4})$-load-balancing sparsifiers, and ChatGPT gave a short counterexample. We then generalized and simplified this bound to obtain \Cref{thm:linear-size-lower} and its proof. 

\bibliographystyle{alpha}
\bibliography{refs}

\end{document}